\documentclass[useAMS,usenatbib,usedcolumn,referee]{mnras}
\usepackage{graphicx,multicol}
\usepackage{times}

\usepackage{slashbox}
\usepackage[T1]{fontenc}
\usepackage[utf8]{inputenc}
\usepackage{amsmath}
\usepackage{amssymb}
\usepackage{epsfig}

\newcommand{\D}{\Delta}
\newcommand{\be}{\begin{equation}}
\newcommand{\ee}{\end{equation}}
\newcommand{\bsube}{\begin{subequations}}
\newcommand{\esube}{\end{subequations}}
\newcommand{\ba}{\begin{array}}
\newcommand{\ea}{\end{array}}

\newcommand{\bea}{\begin{eqnarray}}
\newcommand{\eea}{\end{eqnarray}}
\newcommand{\calN}{{\mathcal N}}
\newcommand{\calL}{{\mathcal L}}
\newcommand{\vhat}{\widehat{v}}
\newcommand{\uhat}{\widehat{u}}
\newcommand{\calLhat}{\widehat{\mathcal L}}
\newcommand{\calNhat}{\widehat{\mathcal N}}
\newcommand{\dt}{\D t}
\newcommand{\dx}{\D x}

\newcommand{\pvint}{\;-\hspace{-0.45cm}\int}
\newcommand{\lcorr}{l_{\rm corr}}
\newcommand{\kcorr}{k_{\rm corr}}

\title[Relativistic charge solitons]{Relativistic charge solitons created
 due to nonlinear Landau damping:
A candidate for explaining coherent radio emission in pulsars}
\author[Lakoba, Mitra \& Melikidze]{Taras Lakoba$^{1}$,
Dipanjan Mitra$^{2,3}$,  George Melikidze$^{3,4}$\\
$^{1}$ Department of Mathematics and Statistics, 
University of Vermont, Burlington VT 05401, USA \\
$^{2}$ National Centre for Radio Astrophysics, 
Tata Institute of Fundamental Research, Pune 411007, India \\
$^{3}$ Janusz Gil Institute of Astronomy, 
University of Zielona G\'ora, ul. Szafrana 2, 65-516 Zielona G\'ora, Poland \\
$^{4}$ Abastumani Astrophysical Observatory, 
Ilia State University, 3-5 Cholokashvili Ave., Tbilisi, 0160, Georgia\\
}

\begin{document}



\maketitle

\label{firstpage}

\begin{abstract}
A potential resolution for the generation of coherent radio emission in
pulsar plasma is the existence of relativistic charge solitons,
which are solutions of nonlinear Schr\"{o}dinger equation (NLSE). 
In an earlier study, Melikidze et al. (2000)
investigated the nature of these charge solitons; however, their
analysis ignored the effect of nonlinear Landau damping, which is inherent 
in the derivation of the NLSE in the
pulsar pair plasma. In this paper we include the effect
of nonlinear Landau damping and obtain solutions of the NLSE
 by applying a suitable numerical scheme.
We find that for reasonable parameters of the cubic nonlinearity and 
nonlinear Landau damping,
soliton-like intense pulses emerge from an initial disordered state of
Langmuir waves and subsequently propagate stably over sufficiently long times,
during which they are capable of exciting the coherent curvature
radiation in pulsars.
We emphasize that this emergence of {\em stable} intense solitons 
from a disordered state does not occur in a purely cubic NLSE;
thus, it is {\em caused} by the nonlinear Landau damping.

\end{abstract}


\section{INTRODUCTION}
\label{intro}

Radio pulsars are rotationally powered neutron stars where
the radio emission arises well within the neutron star magnetosphere.
Observations of pulsar wind nebula suggest 
that the pulsar wind is composed of a dense electron position pair 
plasma outflowing from the pulsar. 
The problem of solving  the pulsar magnetosphere equations
to obtain estimates of the radiation and pulsar wind from a pulsar,
is nontrivial and is a matter of intense research 
(see, e.g., \citealt{2011ASSP...21..139S}; \citealt{2016JPlPh..82e6302P}). 

The region around a strongly magnetized ($\mathbf{B}\sim 10^{12}$ G) and 
fast-spinning neutron star generates enormous electric fields $\mathbf{E}$ 
and cannot be maintained as vacuum (\citealt{1969ApJ...157..869G}). 
Most theories follow the idea that the region around the neutron star
is a charge-separated magnetosphere that is force-free, 
meaning that the electromagnetic
energy is significantly larger than all other inertial, 
pressure, and dissipative forces. 
To maintain co-rotation in the magnetosphere, the condition 
$\mathbf{E}\cdot\mathbf{B}=0$ 
should be satisfied, and this corresponds to a charge number density 
equal to the Goldreich--Julian density $n_{GJ} = \Omega\,B/(2\pi c e)$, 
where $\Omega = 2\pi/P$, $P$ is the rotational period of the pulsar, $c$ is
the velocity of light, and $e$ is the electron charge.
The magnetosphere is initially charge-starved, and a supply of 
charged particles
can come from the neutron star or due to pair creation in 
strong magnetic fields. 
It was first suggested by \citet{1971ApJ...164..529S} that the region above the 
polar cap is the most likely place of 
electron--positron pair generation by magnetic field, 
and an electromagnetic cascading effect can multiply the pairs
to reach density of
 about $10^{4}-10^{5} n_{GJ}$ (see, e.g., \citealt{2015ApJ...810..144T}). 
This value agrees very well with the evidence
available from observations of pulsar wind nebula (see, e.g.,
 \citet{2011ASSP...21..624B} for a recent review). Thus, pair creation
in the polar cap is an essential feature of any pulsar model.

In the last few decades, significant progress has been 
made in understanding the global force-free magnetosphere physics. 
In the presence of a copious supply of pair plasma and for the
commonly assumed dipolar 
magnetic field configuration, the steady state global current distribution, 
the pulsar wind,
and the resultant magnetic field structure  can be found 
numerically. A large number of studies has been devoted to finding the
global magnetospheric structure (e.g. 
\citealt{1999ApJ...511..351C};\citealt{2006ApJ...648L..51S}; 
\citealt{2006MNRAS.368.1055T};\citealt{2009A&A...496..495K}), and hence the 
global current distribution is considered to be known. 
However, most of these global magnetosphere
studies do not include the effect of how the plasma is generated
in the polar cap. To address this shortcoming,
\citet{2013MNRAS.429...20T} 
and \citet{2010MNRAS.408.2092T} revisited two earlier
models by \citet{1979ApJ...231..854A} and \citet{1975ApJ...196...51R},
where charges can and cannot, respectively, 
be extracted from the neutron star surface.
They combined properties of the global force-free 
 magnetosphere and the local mechanism of pair creation
 and obtained the solution for the plasma generation in the polar cap 
numerically.  
Importantly, \citet{2013MNRAS.429...20T} 
and \citet{2010MNRAS.408.2092T}
 found that the plasma flow along the open dipolar 
field lines is non-stationary, as it was suggested 
by \citet{1975ApJ...196...51R}.

Radio emission from pulsar is thought to arise from the development
of plasma instabilities in the electron--positron plasma 
streaming relativistically along open dipolar magnetic field lines 
in the pulsar magnetosphere.  
However, identifying the physical process that can explain the 
radio emission {\em properties}
in pulsars is a challenging problem in astrophysics. The key issues here
are: \ (i) \ to explain the problem of coherency, which manifests itself
as observed
pulsar radio emission with unrealistically high brightness temperatures 
 $\sim 10^{28}\ldots\;10^{30}$K;
and \ (ii) \ to explain the range of pulsar phenomena, 
such as micropulses, subpulse
drift, nulling/moding, pulsar profile stability, polarization properties, etc..
Generally, the coherent pulsar radio emission
can be generated by means of either a maser or a coherent
curvature mechanism (e.g., \citealt{1975ARA&A..13..511G};
\citealt{1975ApJ...196...51R}; \citealt{1980Ap.....16..104M}; 
\citealt{1984Ap.....20..100M}; \citealt{1991MNRAS.253..377K}; 
\citealt{2000ApJ...544.1081M})
emitted in strongly magnetized
electron--positron plasma well inside the light cylinder.
However, as we will discuss in section~\ref{sec2} (see also
 \citet{2017JApA...38...52M} for a recent review), 
a large body of observations appear to suggest that
the pulsar radio emission is excited via a mechanism of 
coherent curvature radiation.
This radiation emerges from regions of about 500 km above the 
neutron star surface. 
The high brightness temperature
 of this coherent radiation can be explained only if it is excited by
charge bunches containing a very large number of charged particles rather than 
by a single charge.  
The physics of how these charge bunches are formed and how they
emit coherent radio emission is still poorly understood.
In this work we will focus on the problem of 
formation of charge bunches and their stability, and will also
address the problem of coherency in pulsar radio emission.

We will rely on a commonly used approximation whereby 
the non-stationary flow of the plasma along open dipolar field lines
is one-dimensional. This approximation is justified because of the
strong confinement of the plasma along those lines. 
We also note that the recent
time-dependent model of \citet{2010MNRAS.408.2092T}
qualitatively reproduces the non-stationary plasma flow that was
proposed in the classical radio pulsar emission model of 
\citet[hereafter RS75]{1975ApJ...196...51R}.
While the RS75 model does not solve for the detailed time-dependent effect 
of pair creation, it does give a prescription of how to estimate 
the plasma parameters
in the radio emission region. Since we are primarily interested in 
simple estimates of the plasma parameters, 
we will use the RS75 model as the starting point of our study.

RS75 were amongst the first to propose a
model that attempted to explain the overall aspect
of the pulsar emission, i.e., both coherency and radio pulsar observational 
phenomenology. In their model, there exists an inner acceleration region close
to the polar cap, where a relativistic non-stationary flow of 
the electron--positron pair plasma can be established.
To address the problem of 
coherent radio emission, RS75 suggested that charge bunches could be formed due to 
development of a two-stream instability that results from 
the overlap between fast-moving 
and slow-moving particles of the non-stationary plasma.
This instability leads to the formation of linear electrostatic 
Langmuir waves, whose frequency is the plasma frequency. 
As the Langmuir wave propagates along the magnetic field, each type of particles
is subject to the sinusoidal electric field, where for half of its period
the field bunches together charges of one sign, while
for the next half-period it bunches together charges of the opposite sign. 
RS75 proposed that these charge bunches can excite the coherent radio
emission.

However, the explanation of coherent emission as occurring from such 
charge bunches has the following fundamental difficulty, as was pointed out by
\citet{1986FizPl..12.1233L} and \citet{2000ApJ...544.1081M} (hereafter MGP00). 
On one hand, the spatial dimension $\Lambda_b$ of an emitting bunch
(along the magnetic field lines) 
should be smaller than the period of the coherently emitted wave $\lambda_c$:
\bsube
\be
\lambda_c > \Lambda_b.
\label{contr1a}
\ee
Indeed, if $\lambda_c<\Lambda_b$, then different regions of the bunch would 
emit independently and hence incoherently. 
As described above, the bunching is caused by linear Langmuir waves (having
wavelength $\lambda_l$), and the size of a bunch is about half of the wave's
period; i.e., $\Lambda_b\approx \lambda_l/2$. 
Since Langmuir waves have an approximately vacuum dispersion relation,
$\omega = 2\pi c/\lambda$, the condition \eqref{contr1a} that the 
emission be coherent amounts to 
\be
\omega_c < 2\omega_l,
\label{contr1b}
\ee
\label{contr1}
\esube
where $\omega_c$ and $\omega_l$ the characteristic frequency of the emitted waves
and the Langmuir waves, respectively. 
On the other hand, the temporal period of the emitted wave, 
i.e. $\mathcal{T}_c=2\pi/\omega_c$,
 cannot exceed the time window over which the emitting bunch exists;
this time window is half of the period of the Langmuir wave, 
i.e. $\mathcal{T}_b=\pi/\omega_l$.
Indeed, if the condition
\bsube
\be
\mathcal{T}_c  < \mathcal{T}_b
\label{contr2a}
\ee
does not hold, the charge bunch would disperse away before 
it has the chance to emit a radio wave. 
Equivalently to \eqref{contr2a}, one must have
\be
\omega_c > 2\omega_l\,.
\label{contr2b}
\ee
\label{contr2}
\esube
Clearly, the above two conditions: \eqref{contr1b} (coherency of the emission) 
and \eqref{contr2b} (non-dispersal of the charge bunch) are in contradiction
with each other.

In the last few decades, significant refinement of the basic physical ideas
that were postulated by RS75 has been achieved both theoretically and observationally 
(e.g.,  MGP00; \citealt{2004ApJ...600..872G}; 
\citealt{2009ApJ...696L.141M}; \citealt{2014ApJ...794..105M}). 
To circumvent the fundamental difficulty described in the previous paragraph,
MGP00 accounted for nonlinear effects due to sufficiently strong
two-stream instability in the relativistic plasma. Their theory led to
the nonlinear Schr\"{o}dinger equation (NLSE) with a
nonlinear Landau damping term, which describes propagation of the
{\em slowly varying envelope} of Langmuir waves. 
It is important to clarify that the same mechanism --- the interaction
between packets of Langmuir waves and charged particles in the plasma ---
leads to the appearance of both the local and nonlocal nonlinear
terms in the NLSE (see section 4 for details). 
Therefore, strictly speaking, both these terms are to be kept in a
comprehensive analysis of the problem. However, no analytical solution
of the NLSE with the nonlocal nonlinear Landau damping term is known.
Thus, by way of approximation, 
MGP00 neglected the nonlinear Landau damping term, assuming it to be small,
and showed that for reasonable pulsar parameters, the solution of 
the NLSE leads to formation of a nonlinear solitary wave, 
i.e., a soliton, which carries an effective charge.
Unlike the ``half-period" charge bunches in the linear
RS75 theory, the charge solitons can exist for times much longer than 
$\pi/\omega_l$. Thus, 
since $\mathcal{T}_b$ is no longer related to $\pi/\omega_l$, 
condition \eqref{contr2b}
can no longer be deduced from condition \eqref{contr2a}.
(Let us note, in passing, that for solitons, 
condition \eqref{contr1b} also does not follow from condition \eqref{contr1a},
because the soliton's length is much greater than the spatial period of the carrier
Langmuir wave.)
Hence, the bunch non-dispersal condition \eqref{contr1b} no longer 
 contradicts the coherency condition \eqref{contr1a}, and therefore
charge solitons, at least in principle,
can excite coherent radio emission in the plasma.

Yet, an explanation of the coherent emission relying on solitons of
the ``pure" NLSE without a nonlinear Landau damping term
has a shortcoming of its own. A stably propagating soliton (or a few solitons)
is known to emerge only from a certain class of initial conditions --- a 
localized one. However, there is no reason to assume that such an initial
state actually occurs in a magnetospheric plasma; rather, the initial condition
there is likely to be 
a nonlocalized Langmuir wave with a randomly modulated 
envelope.\footnote{
As we discuss in detail in section 5, the random variation of the field's
envelope occurs over a spatial scale that is much larger than the Langmuir
period.
  }
A solution developing from such an initial condition is known to
be a disordered ensemble of solitons and a non-solitonic part of
the solution (so-called linear dispersive waves). In this
disordered state, solitons continuously appear and disappear as a result
of their interaction with one another and with linear dispersive waves; 
see, e.g., \citet{2007Natur.450.1054S}; \citet{2010PhFl...22c6601F};
\citet{2015PhLA..379.1821L}; \citet{15_AgafontsevZakharov}; 
\citet{18_GelashAgafontsev}.
Consequently, such ``flickering" solitons do not exist for times long enough that
would let conditions \eqref{contr2a} and \eqref{contr1a} hold
simultaneously. Thus, a mechanism that would preserve a soliton's
individuality for a sufficiently long time,
is required for the MGP00 theory to become a strong contender in explaining
the pulsar coherent radio emission.

In this paper we demonstrate that taking into account the effect of 
nonlinear Landau damping
in the MGP00 theory provides such a soliton-stabilizing mechanism. 
The main part of this paper is organized as follows. 
In section~\ref{sec2} we briefly describe the observational evidence from radio pulsars 
that motivates invoking the charge soliton model. In section~\ref{sec3} 
we briefly outline the generation mechanism and features of the radio emitting 
plasma based on the polar-cap RS75 class of models. In section~\ref{sec4} we 
introduce the concept of the NLSE in pulsar plasma, and in section ~\ref{sec5}
we discuss the range of parameters which are reasonable to expect in 
charge bunches of plasma near a pulsar. In section~\ref{numsim} we present
the main results: a numerical observation of an intense long-living 
electrostatic pulse with an internal structure, which is formed in 
the NLSE model 
{\em due to} 
the nonlinear Landau damping. 
In section~\ref{sec6} we summarize
the results.  Appendix A contains a description of the numerical method,
Appendix B discusses the appropriateness of using periodic boundary conditions
in the numerical simulations,
and Appendix C lists definitions of notations used in this work.


\section{OBSERVATIONAL EVIDENCE OF COHERENT CURVATURE RADIATION OF PULSARS} 
\label{sec2}

Radio pulsar phenomenological studies performed over the years provide 
a sound basis for understanding some general properties of the pulsar
radio emission (see, e.g., \citealt{2017JApA...38...52M}). Pulsars emit
periodic signals with period ranging from about 1 msec to 
8.5 sec, and the pulsed emission is restricted
to an emission window which is typically 10\% of the pulse period. In this
study we will focus on properties of so-called normal pulsars, whose 
periods, $P$, are longer than $\sim$50 msec and whose surface 
dipolar magnetic field is about $\sim10^{12}$ G. In normal pulsars the 
average pulse profile, which is obtained by averaging a large number 
of single pulses, is seen to be highly structured and can consists of one to 
several Gaussian-like components.

Pulsars are also highly linearly polarized,
and the polarization position angle (PPA) across the pulsar profile
shows a characteristic S-shaped swing. This has been interpreted
by the rotating vector model (\citealt{1969ApL.....3..225R}) 
as a signature 
of emission arising due to curvature radiation from charge 
bunches moving along the open dipolar magnetic field lines.
The steepest gradient (SG) point of the PPA traverse corresponds to the
fiducial magnetic plane which contains the rotation and magnetic axes.

Pulsar profile along with linear polarization information 
is used in a statistical sense to infer that the pulsar radio emission
beam is composed of a central core emission surrounded by nested conal emission.
The components in the single pulses are more dynamic in their location
inside the pulse window, which leads to such phenomena as: \  
(i) \ subpulse drifting,
where in subsequent single pulses the emission components
are seen to systematically move across the pulse window; \ 
(ii) \ the small-scale quasiperiodic temporal structures seen in 
components of single pulses called "micro-structures"; and \ (iii) \ 
nulling and moding, where
the average or radio emission either switches off completely or changes its pattern
for a certain duration, and then returns back to its original state. All these
phenomena can be considered as non-stationary effects in the 
pulsar magnetospheric plasma.  

In the following three subsections
 we will briefly summarize the basic outcome from pulsar radio
observations and point out the constraints they provide in formulation
of the theory of coherent radio emission from pulsar.


\subsection{Emission height} 

There are three 
different techniques that can be used to determine 
the location where the radio emission detaches from the pulsar magnetosphere. 
Two of these techniques, namely the geometrical method and the aberration and retardation 
(A/R) method,
rely on the fact that pulsar emission arises
in the region of open dipolar diverging magnetic
field lines; merits, drawbacks, and usage of these height estimation methods can 
be found in \citet{2004A&A...421..215M} and \citet{2008MNRAS.391..859D}. 
Between these two methods, the A/R method, 
proposed by \citet{1991ApJ...370..643B}, is known to give
more robust estimates for radio emission locations in normal pulsars.
   \footnote{ 
	The geometrical method involves estimation of emission heights 
by solving for the geometry of the pulsar beam. The solution,  in turn,
involves fitting the rotating vector model to the PPA traverse to estimate 
the angle between the rotation axis and magnetic axis as well as
the angle between the magnetic axis and the observer's line of sight. 
These estimated parameters turn out to be highly correlated (see e.g.
\citet{2001ApJ...553..341E}), and hence robust estimates of actual 
height using this method are not possible.
   }
Moreover, the A/R method revealed that emission heights can be 
estimated independently of pulsar's geometry (see \citealt{2004ApJ...607..939D}). 
The A/R effect is seen 
as a shift between the
center of the total intensity profile and the fiducial plane containing the magnetic and
spin axes, which is often identified as the steepest gradient point of the PPA traverse or
the peak of the core emission. The A/R methods suggest that the 
core and conal emission, i.e. the overall emission across the pulsar
beam, arises from approximately the same height (\citealt{2016MNRAS.460.3063M}). 
A few notable studies dedicated to finding emission heights using 
the A/R method are: 
\citet{1991ApJ...370..643B},
\citet{1997A&A...324..981V}, 
\citet{2004A&A...421..215M},
\citet{2011ApJ...727...92M},
\citet{2008MNRAS.391.1210W}.
These studies suggest that the radio emission arises from about 
$\sim$500 km above the neutron star’s surface (see also Fig.~3 of 
\citealt{2017JApA...38...52M}) 
The third method for finding emission heights is based on using pulsar scintillation. 
In this method,
one uses the fact that the emission from the compact emission region of the pulsar
passes through the interstellar medium which can act as a varying lens, thus modulating
the pulsar signal. The nature of this modulation depends on the spatial transverse 
extent of the source, 
which can be recovered by performing extremely high spatial resolution interferometry. 
The method has been applied successfully on a few pulsars, and accurate results are 
only available for the Vela pulsar which imply that the spatial transverse extent of
the emission source is about 4 km and the corresponding radio emission altitude is 
estimated to be about 340 km, in agreement with the other methods 
(\citealt{2012ApJ...758....8J}).

The pulsar radio emission height $R_{em} \sim 500$ km  is a very significant input to 
the pulsar radio emission mechanism problem. 
The only plasma instability that can develop
 at these heights (where the magnetic field is very strong
and the plasma is constrained to move along the magnetic field lines) 
is the two–stream instability.
Hence, resonance-type instabilities like the cyclotron maser 
instability (which can develop only
near the light cylinder, where the magnetic field is weak), can be ruled out.


\subsection{Evidence for curvature radiation}  

The estimated emission heights $R_{em}$ is the location where the emission
detaches from the pulsar magnetosphere. It is quite possible that the pulsar
emission is generated in the emitting plasma at a certain height $R_g < R_{em}$, and then emerges
out of the plasma at $R_{em}$. There is, however, no direct way to probe this effect,
and one has to resort to wave propagation properties in 
electron--positron plasma at such strong magnetic fields. Once the radiation is generated
in the plasma, it naturally splits as ordinary, or O-mode (polarized 
in the plane of the wave vector $\mathbf{k}$ and the magnetic field 
$\mathbf{B}$) and 
the extraordinary, X-mode (polarized perpendicularly to the $\mathbf{k}$ 
and $\mathbf{B}$ plane). The O-mode
strongly interacts with the plasma and is ducted along the magnetic field lines
or can be damped, while the
X-mode can escape the plasma at $R_g \sim R_{em}$ as if
it were in vacuum (see \citealt{2004ApJ...600..872G}, \citealt{2009ApJ...696L.141M},
and \citealt{2014ApJ...794..105M} for details).

It turns out that there is multiple observational evidence that 
allows determination of the orientation of the emerging polarization
direction with respect to the dipolar magnetic field planes. The most direct
evidence comes from the 
x-ray image of the Vela pulsar wind nebula and fiducial or the SG point 
of the absolute PPA, 
which can be used to establish that the electric vector emanating out of the pulsar 
is orthogonal to the
magnetic field planes, and hence represents the extraordinary (X) mode. 
\citet{2001ApJ...549.1111L} also showed that the proper motion direction (PM) of the pulsar
is aligned with the rotation axis. \citet{2005MNRAS.364.1397J} and
 \citet{2007ApJ...664..443R} produced
a distribution of the quantity $|{\rm PM}- \mbox{absolute PPA}|$ for a few pulsars and found a bimodal distribution
around zero and 90$^{\circ}$. Assuming that the pulsar's PMs are parallel to the rotation
axis, the bimodality could be explained as 
occurring due to the emerging radiation being either parallel or
perpendicular to the magnetic field planes, since pulsars are known to have orthogonal
polarization modes. Alternatively, PMs of pulsars can also be parallel or perpendicular
to the rotation axis. While both the above explanations are possible, it is clear that the
electric vectors of the waves which detach from the pulsar magnetosphere to reach the
observer follow the magnetic field planes. 

These observations can hence be interpreted as suggesting that the observed emission
is associated with curvature radiation mechanism, since this is the only known
emission mechanism that can distinguish the magnetic field planes. 
Further evidence of curvature radiation can also be obtained from single
pulse polarization, where \citet{2009ApJ...696L.141M} 
demonstrated that the instantaneous polarization of components of single pulses
closely follow the average PPA. 


\subsection{Evidence for non-dipolar surface magnetic field}

At a distance of $R_{em}$ where the pulsar
radio emission originates, the magnetic field  is significantly dipolar. However, 
the magnetic field at the surface of the neutron star needs to be significantly 
non-dipolar, so that a sufficient amount of the electron--positron 
pair plasma can be generated to explain the observed pulsar
radiation. Pulsars are known to slow down at a certain rate $\dot{P}$, and 
this slow-down can be used to estimate only the surface dipolar magnetic field 
component to be $B_d \sim 6 \times 10^{19} \sqrt{P\dot{P}}$ G;
 here $P$ is the pulsar rotation period (in seconds) and $\dot{P}$ is 
non-dimensional.
There is, however, observational evidence that suggests the 
presence of a surface non-dipolar magnetic field. 
The strongest piece of such evidence, from which the existence of non-dipolar 
magnetic field can be inferred, 
 comes from the discovery of a long-period ($P=8.5$ sec)
pulsar PSR J2144$-$3933 (\citealt{1999Natur.400..848Y}). 
\citet{2001ApJ...550..383G} argued 
that significant creation of pair plasma in this pulsar, which 
is essential for producing the 
radio emission, can only happen if the radius of 
curvature of the surface magnetic field is $\rho \sim 10^5$ cm, 
which is about an order of magnitude smaller than $\rho$ values
in normal pulsars with $P\sim 1$.
The smaller value of $\rho_c$ in PSR J2144$-$3933 
implies that the magnitude of the 
surface non-dipolar magnetic field there is about 10$^{14}$ G, 
which is about 100 times higher than the dipolar magnetic field. 
Furthermore,  in some radio pulsars, soft x-ray blackbody
radiation is seen from hot polar caps, and the estimated area of the polar cap is often 
found to be smaller than the dipolar area, suggesting the presence of a strong
non-dipolar field on the neutron star surface (see, e.g., Table 1 of 
\citet{2017JApA...38...46G} and references therein). 

Thus, in summary, the basic input to the pulsar emission models from observations
is that coherent radio emission is excited in a non-stationary plasma 
flowing away from a pulsar and detaches from it at a height of about 
a few hundred km above the pulsar's surface,
 in a region of open dipolar magnetic field lines. The magnetic field
on the neutron star surface, however, is significantly non-dipolar.

\section{Plasma condition in the magnetosphere and Mechanism for Radio Emission in Pulsars}
\label{sec3}

As we have described in the Introduction,
the RS75 class of the polar cap models provide a framework whereby 
the observed coherent curvature radiation is attributed to
the emission of radio waves by charge bunches in the plasma. 
In this section we will summarize subsequent stages of creation
of this radio emission. We will refer to 
a pulsar, i.e., a neutron star, having the following parameters:
radius $R_s$, pulsar period $P$ (measured in seconds),
pulsar slow-down rate $\dot{P}$, surface 
magnetic field $B_s$, and dipolar magnetic field $B_d$
(see the beginning of section \ref{sec2}.3). 
For future use we also introduce the
ratio $b=B_s/B_d$, a non-dimensional parameter 
$\dot{P}_{-15} = \dot{P}/(10^{-15})$, which is assumed to be of order one, 
and the vector of angular velocity of the rotating star, whose magnitude
is $\Omega=2\pi/P$.

\subsection{Gap formation}

RS75 suggested that if the condition 
$\mathbf{\Omega\cdot B_s} < 0$ holds above the pulsar polar cap, then the polar
cap is positively charged. They envisaged a situation where 
initially there is only a limited supply of stray positive charges above 
the polar cap, which is relativistically flowing away from the pulsar 
along open magnetic field lines as a pulsar wind. Consequently, if 
the binding energy of the 
ions in the neutron star surface is sufficiently strong, then the region 
above the polar cap will be deficient in positive charges, and
a vacuum gap can be created, where an enormously 
high electric field exists. 
Photons of energy $> 2 m_e c^2$, where $m_e$ is the mass of electron,
are split inside the vacuum gap into electron--positron 
pairs, and the electric field in the gap separates these two types of charges.
They are then further accelerated along the curved magnetic field lines
(hence the term `curvature radiation')
and can generate high-energy curvature photons,
which, in their turn, after traveling some mean free path,  
can produce other electron--positron pairs.
In terms of pulsar parameters, the potential drop $\Delta V$ across the gap and 
the gap's height $h$  can be expressed as
\begin{equation}
\Delta V \sim 2 \times 10^{12} b^{-1/7} P^{-3/14} \dot{P}_{-15}^{-1/14} \rho_6^{4/7}~~\rm{V} ,
\label{deltaV}
\end{equation}
\begin{equation}
h \sim 5 \times 10^3 b^{-4/7} P^{1/7} \dot{P}_{-15}^{-2/7} \rho_6^{2/7}~~\rm{cm} 
 \, \gtrsim 10^4 ~\mathrm{cm}\,.
\label{height}
\end{equation}
Here $\rho_6 \equiv \rho\,({\rm cm})\,\times \,10^{-6}$, where 
$\rho$ is the curvature radius of magnetic field lines in the gap region. 
To obtain the numeric estimate in \eqref{height}, we used the values
$\rho_6\sim P\sim \dot{P}_{-15} \sim 1$ and $b\sim 10$.


\subsection{Spark formation} 

A number of such localized discharges 
can be formed in the gap, and each such a discharge causes
a pair creation cascade. The electric field in the gap accelerates 
the electrons towards
the stellar surface, while the positrons are accelerated away from the surface.
At the top of the gap these positrons acquire Lorentz factors of $\gamma_b$  
such that
\begin{equation}
\gamma_b \approx e \Delta V/m_e c^2\,\sim\, 2\times 10^6. 
\label{gammap}
\end{equation}

As these particles move away from the gap 
to the region where $\mathbf{E\cdot B} = 0$, 
they continue to create high energy photons, which further create pairs, 
and this cascade leads to the generation
of a cloud of secondary electron--positron plasma, which has a significantly
lower Lorentz factor with a mean value of $\gamma_p$. 
If the positron number density in the primary beam is $n_b$, then the number 
density of  pairs in the secondary plasma
can be estimated as $n_p \sim (0.5\gamma_b/\gamma_p) n_b$, and thus 
the density of the secondary plasma increases by a 
factor $\kappa = n_p/n_b$. 
In this work we will use the value $\kappa\sim 10^4$ (\citealt{71_Sturrock}).
The burst of pair production process
increases the charge density along the gap discharge stream and screens 
the potential in the gap. This process develops exponentially, 
and after a certain distance from the star, which is estimated to be 
$\sim 30-40 h \sim 500~{\rm m}$ (RS75), the charge density becomes close to $n_{GJ}$, 
and the particle acceleration process stops.  During this time 
the discharge spreads in the lateral direction thus acquiring a width of $\sim h$.
 We will call this fully formed discharge a spark, and each
spark is associated with a secondary plasma cloud.
According to the above description, 
such a cloud has the shape of a column with a longitudinal dimension 
of $\sim 500~{\rm m}$ and a diameter of $\gtrsim 10~{\rm m}$ near the
pulsar's surface. 
As the cloud moves away from the star to a distance $R$ from its center, 
the cloud's diameter grows proportionally
to $(R/R_s)^{1.5}$ due to the divergence of dipolar magnetic field lines.

The charge number density of the primary beam, $n_b=n_{GJ}$, 
can be expressed in terms of pulsar parameters as  
\begin{equation}
n_{GJ} \sim 6 \times 10^5 \left(\dot{P}_{-15}/P\right)^{0.5} R^{-3}_{50} \rm{cm^{-3}},
\label{ngj}
\end{equation}
where $R_{50}=R/(50R_s)$; and a value of $R_s = 10$ km will be used for
subsequent  calculations in this paper. 
Here and below we normalize the distance to $50R_s$ since the coherent radio
emission occurs around that altitude (see next subsection); hence $R_{50}\sim 1$. 
The number density of the secondary plasma is $n_p = \kappa n_{GJ}$, and hence
the mean Lorentz factor of the secondary plasma can be estimated to be
\begin{equation}
\gamma_p \approx \gamma_b /(2 \kappa) \sim 100 \,.
\label{gammas}
\end{equation}
The plasma frequency $\omega_p$ is given by
\begin{equation}
\omega_p \sim 4
 \times 10^{9} R_{50}^{-1.5} \kappa_4^{0.5} (\dot{P}_{-15}/P)^{0.25}~ 
  \rm{s}^{-1} , 
\label{omegap}
\end{equation}
where $\kappa_4=\kappa/10^4\,\sim\,1$.

For later use in the next subsection, we will also mention that 
once the spark-associated plasma cloud leaves the polar cap 
and the gap's potential can no longer be screened,     
the next discharge can be initiated. Thus, the distance between two consecutive
plasma clouds is estimated to be $\sim h$.

\subsection{Development of linear two-stream instability in secondary plasma}

The overall sparking process leads to a non-stationary flow of successive 
plasma clouds flowing along a bundle of magnetic dipolar field lines. 
Indeed, since the magnetic field is strong and the ratio of the plasma frequency 
to cyclotron frequency $\omega_p/\omega_B \ll 1$ at the radio emission 
heights, the charges are confined to move tightly along the magnetic field
lines. Therefore, we will hence restrict ourselves to discussing the 
plasma properties in a one-dimensional flow.

It should be mentioned that within each plasma cloud, distribution functions 
of electrons and positrons differ from each other (see the discussion around 
Eq.~\eqref{deltagamma}
below). RS75 pointed out that it could be a reason for the two-stream instability. 
On the other hand, \citealt{87_Usov} suggested that due to an overlap of the slow- and 
fast- moving particles of two {\em successive} clouds, a two-stream instability can develop. 
\citet{1998MNRAS.301...59A} showed that this 
two-stream instability provides a sufficiently strong Langmuir turbulence in 
the plasma, with the frequency of Langmuir waves $\omega_l$ being
\begin{equation}
\omega_l \approx  \gamma_p \omega_p \sim 4 \times 10^{11} R_{50}^{-1.5} \,.
\label{omegal}
\end{equation}
Moreover, disturbances of the envelope of these slowly modulated Langmuir waves,
obeying the NLSE (MGP00),
propagate with the group velocity that can coincide with the velocity of some
portion of the charged particles. This occurs because of the aforementioned 
spread of particle velocities, both within one species and well as between the
two species. This synchronism is behind the mechanism of the nonlinear Landau 
damping, first derived for this situation by MGP00. To quantify that effect,
as well as the magnitude of nonlinear and dispersive terms in the NLSE in the
next section, we will need to refer to the velocity distribution functions
of electrons and positrons in the secondary plasma. They can be approximated
by a Gaussian function centered around momentum $p_{\alpha}$ and 
having a spread of $p_{T}$:
\be
f_{\alpha} \propto \exp[-((p-p_{\alpha})/p_T)^2]  
 \sim \exp[-((\gamma - \gamma_{s,\,\alpha})/\gamma_T)^2].
\label{distrf}
\ee
Here $\alpha = +$ or $-$, corresponding to 
positron or electron species of the plasma, and $\gamma_{s,\,\alpha}$ and
$\gamma_{T}$ are the Lorentz factors corresponding to the momenta $p_{\alpha}$
 and $p_T$. 
It is important to notice that due to the flow of plasma along the 
curved magnetic field lines, the distribution functions of positrons and electrons 
become unequal: $f_{+} \neq f_{-}$ (\citealt{1979ApJ...229..348C}).
It was further shown by \citet{1998MNRAS.301...59A} that
\begin{equation}
|\gamma_{s,\,+}-\gamma_{s,\,-}|/\gamma_p \equiv 
\Delta \gamma/\gamma_p \approx \Delta \sigma \gamma_p^3/\gamma_b \,,
\label{deltagamma}
\end{equation}
where the value of parameter $\Delta \sigma$ depends on whether the surface
magnetic field at polar cap region is assumed to be strictly dipolar or 
to have a multipolar structure (MGP00); see section 2.3. Importantly for the estimate of
the nonlinear Landau damping parameter is section 5 below, it was shown in MGP00
that $\Delta \gamma/\gamma_p$ is in the range $0.5\ldots 2$.

Finally, an important parameter when solving the NLSE in subsequent sections
is the distance from the pulsar's surface where the two-stream instability
develops, thereby leading to strong Langmuir turbulence and hence strong
radio emission. 
Simple kinematic estimates performed by MGP00 show that the distance at which this
instability can set in is $\sim 2\gamma_p^2 h \gtrsim 200$ km. 
In terms of pulsar parameters, this can be written as
\bsube
\begin{equation}
R_{\rm onset} \sim 20R_s\times \gamma_2^2 \rho_6^{2/7} b^{-4/7} B_{12}^{-4/7} P^{3/7}\,,
\label{rin}
\end{equation}
where $B_{12}=B_d/ 10^{12}~~{\rm{G}}$ and $\gamma_2 = \gamma_p /10^{2}$.
The longitudinal dimension of the instability region is limited by the divergence
of the magnetic field lines, which leads to a decrease of the plasma charge density
in proportion to $(R/R_s)^{1.5}$. 
As suggested in \citet{1998MNRAS.301...59A},
 the dimension of this region is 
\begin{equation}
\Delta R \sim 500 ~~{\rm km}. 
\label{rinl}
\end{equation}
Therefore, we can estimate that most of the radiation comes from the middle (or the 
second half) of this region, where, on one hand, the Langmuir turbulence is 
{\em already} strong because the two-stream instability has had sufficient time to
develop, and, on the other hand, is {\em still} strong enough, having not been 
weakened by the divergence of the magnetic field lines. Thus, the location of
the radiating region can be estimated to be
\be
R_{em} \sim (R_{\rm onset}+\Delta R/2) \sim 500~~{\rm km}
\label{Rrad}
\ee
\esube
from the star. Equivalently, one has $R_{50}\sim 1$ in this region. This is the 
value we used in \eqref{omegap} to obtain the estimate \eqref{omegal}. 
Let us note that estimate \eqref{Rrad} is in good agreement with observations,
as discussed in section 2.1.

In the next two sections we will discuss the nonlinear equation satisfied by
the envelope of the Langmuir waves and the range of parameters for which
it should be solved. Then, in section 6, we will demonstrate that the solution
of that equation is capable of explaining features of the observed coherent 
radiation from pulsars. 


\section{Nonlinear dynamics and creation of relativistic charge solitons}
\label{sec4}

The nonlinear interaction of an electric field with plasma,
which involves the nonlinear Landau damping as an integral part,
has been studied for a long time. In one of the early studies,
\citet{73_IchikawaTanuiti} derived the 
NLSE with the nonlinear Landau damping term
for the non-relativistic electron--ion plasma.
This equation has the form:
\be
i \frac{\partial E_{\|} }{\partial \tau} + 
 G \frac{\partial^2 E_{\|} }{\partial \chi^2} + q  E_{\|} 
\left( \big| E_{\|}\big|^2 + 
  \frac{s}{\pi q} \pvint \frac{ | E_{\|}(\chi',\tau)|^2 \; d\chi'}{\chi-\chi'} \;\right)\,
 = \, 0.
\label{NLSwdamp}
\ee
where $E_{\|}$ is the amplitude of the electric field and $\tau$ and $\chi$
are the space and time coordinates. More details about these variables
and the coefficients in \eqref{NLSwdamp} will be given when we discuss
its application to the pulsar plasma. Nonlinear Landau damping is

Later on, \citet{1980Ap&SS..68...61P} and \citet{1980Ap.....16..104M}
showed that the evolution of the envelope of Langmuir wave packets
is also governed by the NLSE \eqref{NLSwdamp} 
for the relativistic electron--ion and electron--positron plasmas,
respectively. 
Here, as well as for other types of plasmas, 
the interaction between the
charged particles of the plasma and those packets of Langmuir
waves whose group velocity coincides with the velocity of the particles,
leads both to the purely cubic, local term in the NLSE and the
nonlocal, nonlinear Landau damping term. 
Let us also note that the NLSE of the same form \eqref{NLSwdamp}
has been derived for other types of plasmas;
see, e.g., \citet{09_NLS_Dusty_NonlinLD}; 
\citet{15_NLS_CollisionlessElectronPositron_NonlinLD}; 
\citet{17_NLS_Degenerate_NonlinLD}; \citet{18_NLS_RelatElectronIon_NonlinLD};
to name just a few studies. 
Related plasma models that also account for nonlinear Landau damping
can be found, e.g., in 
\citet{83_KdV_IonAcoustic_NonlinLD}; 
\citet{88_KdV_RelatElectronAcoustic_NonlinLD};
\citet{89_DNLS_MHD_NonlinLD};
\citet{15_mKdV_DustyBiIon_NonlinLD}; 
\citet{17_KdV_IonAcousticSemiclassical_NonlinLD}.

MGP00 applied the NLSE \eqref{NLSwdamp} to the pulsar plasma. 
In that context, $E_{\|}$ in \eqref{NLSwdamp} 
is the amplitude of the electric field which is parallel to the 
external magnetic field and $\chi$ and $\tau$ are the dimensional space 
and time coordinates in the moving 
frame of reference (MFR). They are related to the coordinates
$\chi_{\circ}$ and $\tau_{\circ}$ in the observer frame of reference (OFR) by: \
$\chi = \gamma_{\circ} (\chi_{\circ} - v_g \tau_{\circ})$ and 
$\tau = \gamma_{\circ}(\tau_{\circ} - (v_g/c^2) \chi_{\circ})$, 
where $v_g$ is the group velocity of Langmuir waves and
$\gamma_{\circ}\approx\gamma_p$  is the corresponding Lorentz factor. 
In particular, any length and time intervals in the OFR are related to such 
intervals in the MFR by: \ $\delta \chi_{\circ} = \delta \chi/\gamma_p$ 
and $\delta \tau_{\circ} = \delta \tau/\gamma_p$. 
The coefficients $G$, $q$ and $s$ in Eq.~\eqref{NLSwdamp} correspond to
the dispersion, cubic nonlinearity, and nonlinear Landau damping terms, respectively.
These coefficients depend on 
the distribution functions $f_{\alpha}(p)$ (see section 3.3) 
and other parameters of the plasma and are given by Eq. (A20) in MGP00: 
\begin{eqnarray}
\label{pareq1}
G & = & \frac{1}{4} \frac{\gamma_p^2 c^2}{\omega_p} G_d\,, \\
\label{pareq2}
q & = & \left( \frac{e^2}{m_e c}\right)^2 \frac{1}{\gamma_p^2 \omega_p} q_d\,, \\
\label{pareq3}
s & = & \left( \frac{e^2}{m_e c} \right)^2 \frac{1}{\gamma_p^2 \omega_p} s_d \,,
\end{eqnarray}
where $G_d$, $q_d$ and $s_d$ depend only on the distribution functions of the 
plasma. 
Typical plots of the coefficient $G_d$ and the ratio $s_d/q_d$ as functions
of the thermal spread $\gamma_T/\gamma_p$ of the plasma (see section 3.3)
are shown in Fig.~\ref{fig_para}.\footnote{
These values were obtained following the procedure described in Appendix A of MGP00,
while correcting an arithmetic error that occurred in that paper.
    } 

\begin{figure}
\begin{center}
\includegraphics[height=5cm,width=6.5cm,angle=0]
{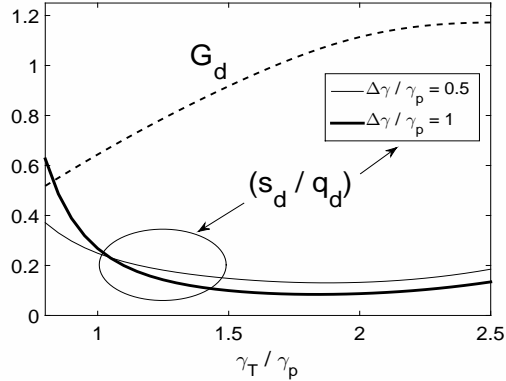}
\caption{Parameters $G_d$ and $s_d / q_d$ in 
\eqref{pareq2} and \eqref{pareq3} as functions of the thermal spread
of the plasma $\gamma_T/\gamma_p$, plotted for two representative values of the 
relative shift $\Delta \gamma / \gamma_p=0.5,\,1$ between the centers of the
electron and positron distributions; see \eqref{deltagamma}. The ellipse 
indicates grouping of the two $s_d/q_d$-curves, to distinguish them from the
$G_d$-curve.}
\label{fig_para}
\end{center}
\end{figure}

Two comments are in order about the coefficients in the NLSE \eqref{NLSwdamp}. 
First, Eqs.~\eqref{pareq2} and \eqref{pareq3} and Fig.~\ref{fig_para}
explicitly demonstrate the point that we have mentioned earlier.
Namely, the same nonlinear process of resonant interaction between 
relativistic charged particles and packets of Langmuir waves
gives rise to both nonlinear terms: the purely cubic one
(proportional to to $q_d$) and the nonlocal one (describing
nonlinear Landau damping and 
proportional to $s_d$). Thus, a comprehensive analysis of the NLSE
\eqref{NLSwdamp} is to take into account both these terms. \ 
Second, it follows from Fig.~1 of MGP00 that 
for the considered ranges of thermal spread $\gamma_T/\gamma_p$ and
relative shift $\Delta\gamma/\gamma_p$ of the positrons' and electrons' 
distribution functions, $q_d$, and hence $q$ in \eqref{NLSwdamp}, is positive. 
Since also $G>0$ (see \eqref{pareq1} and Fig.~\ref{fig_para} above),
then the NLSE \eqref{NLSwdamp} is of the so-called focusing type
($G\cdot q >0$).
It is widely known that in the focusing, purely cubic NLSE, 
solitons can form from an appropriate --- i.e., localized, --- initial state.

MGP00 exhibited and analyzed the soliton solution of the NLSE 
without taking into account the effect of nonlinear Landau damping. 
This approximation had to be made since no analytical solution
of that equation with the nonlinear Landau damping terms is known. 
Physically, this approximation 
is justified only when the nonlinear Landau damping term,
$\propto s$, is so small compared to the nonlinear term, $\propto q$,
that it does not have the chance to affect the evolution of the
Langmuir wave's envelope over typical times considered. 
However, we will show below for reasonable parameters of pulsar plasma,
this assumption does not necessarily hold; hence the effect of nonlinear Landau
damping over such sufficiently long times cannot be neglected.

Our study extends that of MGP00 in two ways. First, obviously, we take
into account the nonlinear Landau damping term by carrying out numerical 
simulations of the NLSE \eqref{NLSwdamp}.
Second, and more importantly, we study the evolution of a 
{\em disordered initial state} instead of a soliton and find a striking
difference between these two evolutions. It should be mentioned that the
approximate evolution of the NLSE soliton under the effect of a small 
nonlinear Landau damping has been considered in a number of studies:
see, e.g., \citet{78_NLS_SolPertNonlinLD};
\citet{89_DNLS_MHD_NonlinLD}; 
\citet{09_NLS_Dusty_NonlinLD}; 
\citet{15_NLS_CollisionlessElectronPositron_NonlinLD}; 
\citet{17_NLS_Degenerate_NonlinLD}; \citet{18_NLS_RelatElectronIon_NonlinLD}.
They all have found that this perturbation leads to the decay and acceleration
of the soliton. {\em In contrast}, we will demonstrate that 
due to the interaction
of the intense pulse (which, as we will argue, is a soliton) with the surrounding
field, the pulse's amplitude will grow to significantly exceed that of the said
field.

With this motivation, we will now proceed to nondimensionalizing
the NLSE \eqref{NLSwdamp} with nonlinear Landau damping,
so as to later solve it numerically. 


\section{Nondimensionalization of Eq.~(13)} 
\label{sec5}

We begin by nondimensionalizing the space variable in the MFR:
\bsube
\be
\chi = l\,\theta\,x,
\label{tn_01a}
\ee
where $x$ is the nondimensional space variable, factor $\theta$ is to be defined
shortly, and the characteristic scale $l$ is defined as:
\be
l= c/\omega_p = \lambda_p/(2\pi)\,,
\label{tn_01b}
\ee
\label{tn_01}
\esube
with $\lambda_p$ being the spatial period of Langmuir waves in the MFR, and 
the plasma frequency $\omega_p$ is given by Eq. \eqref{omegap}. 
The parameter $2\pi\,l$ is of the same order of magnitude as
the Langmuir spatial period $\lambda_p$ (in the MFR) 
at some intermediate location of the 
plasma cloud (see section~\ref{sec3}). Therefore,
factor $\theta/(2\pi)$ in \eqref{tn_01a} 
characterizes the ratio of spatial scales of the Langmuir wave envelope to the Langmuir 
period $\lambda_p$.\footnote{
   In other words, $x=O(1)$ corresponds to
   the dimensional scale of $O\big(\theta/(2\pi)\big)$ of Langmuir 
	 spatial periods in the MFR.}
In what follows we will consider this ratio to be on the order of 
$10^2\ldots 10^3$, whence $\theta=10^3\ldots 10^4$.

We would like to stress that the above assumption about the range for
$\theta$ does not impact the main conclusion of our study, for two reasons.
First, as we will show below, a value of $\theta$ results merely in a
ball-park estimate of the maximum nondimensional simulation time, while
the freedom to adjust the nonlinearity coefficient in the nondimensional equation
will still allow us to observe the important changes in the electric
field's evolution. 
Second, it is realistic to assume that there exists a wider range of
$\theta$ values than what we assumed above; this simply
corresponds to the envelope of the initial electric field having a
wider range of spatial scales. Then, for a given amplitude of the
electric field's envelope (i.e., for a given magnitude of nonlinear 
terms in Eq.~\eqref{NLSwdamp}), only those of its fluctuations whose wavelength fall into 
a narrow(er) range of values will exhibit the phenomenon of pulse
formation described below. In other words, the fluctuations with
the spatial scale of interest to us will be selected by the governing
Eq.~\eqref{NLSwdamp} and not by our assumption of the range
of $\theta$ (as long as that range is sufficiently broad).

In addition to an uncertainty in the value range of our nondimensional 
parameters that occur due to an uncertainty of the scaling parameter $\theta$,
there is also a (much smaller) uncertainty due to these parameters' dependence on the
height $R$ above the pulsar surface where the coherent radiation is emitted.
In \eqref{Rrad} we estimated that this occurs around 500 km above the surface. 
Correspondingly, $R_{50}\sim1$, as defined after Eq.~\eqref{ngj}, and this 
value is to be used to estimate the plasma frequency $\omega_p$ in \eqref{omegap}
and hence the parameter $l$ in \eqref{tn_01b}. For other values of $R_{50}$,
one has $l\propto R_{50}^{1.5}$. We will use this fact in section 6.1 below.

Next, we normalize $E_{\|}$ to a typical magnitude of the electric field,
 $E_0$, at a location where the two-stream instability sets in in the 
cloud of secondary plasma (see \eqref{rin}):
\be
E_{\|} = E_0\,u;
\label{tn_02}
\ee
thus $u$ is the nondimensional electric field. Note that by this definition,
$u=O(1)$ initially.

Finally, we introduce the nondimensional time, $t$, via 
%
\be
\tau = \frac{(l\theta)^2}{G}\,t = \frac{4\theta^2}{\gamma_p^2 \omega_p\,G_d}\; t,
\label{tn_03}
\ee
where we have used formula \eqref{tn_01b} and the relation 
\eqref{pareq1}.
Then, upon dividing through by $G/(l\theta)^2$, 
Eq.~\eqref{NLSwdamp} attains the form:
\be
i \frac{\partial u }{\partial t} + 
  \frac{\partial^2 u }{\partial x^2} + Q\,u 
\left( | u|^2 + 
  \frac{s}{\pi q} \pvint \frac{ | u(x',t)|^2 \; dx'}{x-x'} \, \right)\,
 = \, 0.
\label{tn_04}
\ee
where using Eqs.~\eqref{pareq1} and \eqref{pareq2} we have: \  
$Q = qE_0^2(l\,\theta)^2/G = \big(2E_0\theta\,e^2/(\gamma_p^2 m_e c\omega_p)\big)^2\,q_d/G_d$.
The initial condition to this equation has, by design, the nondimensional
magnitude and spatial scale of order one: $|u|\sim 1$ and $|u|/|u_x| \sim 1$.

Let us now estimate the maximum nondimensional simulation time.
First, the dimensional prototype of that time, $\tau_{\max}$, is that
needed for a Langmuir wave packet to travel, within the cloud of secondary plasma,
a length ${\Delta R}$, where the Langmuir turbulence is sufficiently strong
(i.e. the nonlinear and dispersive terms in \eqref{NLSwdamp} significantly
affect the wave packet's evolution). 
As estimated in Eq.~\eqref{rinl}, 
${\Delta R} \sim 500$ km. 
Since that length is referenced in the OFR while $\tau$ is measured in the 
MFR, the Lorentz factor $\gamma_p\sim 10^2$ (see Eq~\eqref{gammas}) 
needs to be accounted for; thus
\bsube
\be
 \tau_{\max}\approx \frac{\Delta R}{c\,\gamma_p} \sim 10^{-5} {\rm s}.
\label{tn_05a}
\ee
Using now relation \eqref{tn_03}, where $G_d \sim 1$ (see Fig.~\ref{fig_para}), 
and an estimate 
$\omega_p\sim 4\times 10^9$ s$^{-1}$ for a typical pulsar 
with $P \sim \dot{P}_{-15} \sim \kappa_4 \sim R_{50} \sim 1$ 
(see Eq.~\eqref{omegap})), one obtains that:
\be
t_{\max} \sim 10^8/\theta^2 \, \sim \,1\ldots 10^2.
\label{tn_05b}
\ee
\label{tn_05}
\esube
Here the last estimate follows from our 
earlier assumption $\theta=10^3\ldots 10^4$.

Now, as we will demonstrate in the next section, the hallmark of the
evolution of a Langmuir wave packet 
governed by Eq.~\eqref{NLSwdamp} is the formation, out of an
initially disordered state, of an intense pulse
with an internal structure. In light of this, we should consider 
such values of nonlinear parameters $Q$ and $(s/q)$ in the nondimensional
Eq.~\eqref{tn_04} that result in such formation over the times estimated in
\eqref{tn_05b}. As for parameter $(s/q)$, which characterizes 
the strength of nonlinear Landau damping relative to the purely cubic nonlinearity 
(and does not depend on details of nondimensionalization; 
see \eqref{pareq2} and \eqref{pareq3}), its size
can be estimated from Fig.\ref{fig_para}.
For moderately large values of the thermal spread, 
$\gamma_T/\gamma_p\sim 1\ldots 2$,
one finds the size of the nonlinear Landau damping term relative
 to the size of the purely cubic nonlinear term fall in the range
\be
s/q  \,= \, s_d/q_d \,= \,0.05\ldots 0.2.
\label{tn_06}
\ee

On the other hand, it is not possible to estimate the correct order 
of magnitude of the nondimensional parameter $Q$ 
in \eqref{tn_04} based on physical grounds, 
because one does not have an estimate for the unknown electric field $E_0$.
Therefore, we simply had to use trial and error to find a range of values of 
$Q$ such that, for $(s/q)$ being in the range \eqref{tn_06}, the times
of intense pulse formation are in the range \eqref{tn_05b}. As we will
demonstrate in the next section, such $Q$ have the magnitude $O(1)$.

\section{Numerical simulations of Eq.~(20)}
\label{numsim}

In order to solve Eq.~\eqref{tn_04}, we used a numerical method recently proposed in 
\cite{2016JSci..72.14L}. 
This method combines the leap-frog (LF) solver with the idea of the
integrating factor (IF) method and hence will be referred to as IF-LF. For the 
reader's convenience we present its details in Appendix A. The non-physical parameters
of the simulations, such as time step $\dt$ and mesh size $\dx$, are also listed there.

As we mentioned in the Introduction and in section 3, 
the nonlinear evolution of Langmuir waves commences at a distance
of around 200 km from the pulsar; see text before \eqref{rin}.
It is reasonable to assume that the electric field there, amplified
by the two-stream instability, does not have any regular structure but
is, instead, disordered. 
Therefore, we take 
the initial condition $u(x,0)$ for the envelope of the Langmuir wave
in Eq.~\eqref{tn_04}
in the form of a mixture of a constant and a zero-mean random field. 
In the simulations we vary both the ratio $r$ of these 
two parts in the mixture and the correlation length $\lcorr$ of the random part,
since the actual range of these physical parameters in the plasma is not
known. Thus:
\bsube
\be
u(x,0) = (1-r) + r\, \int \frac{\exp[-0.5(k/\kcorr)^2 -ikx]}{\sqrt{\sqrt{\pi}\,\kcorr}} 
\; \widehat{w}(k) \, dk;
\label{t5_01a}
\ee
where $\kcorr=2\pi/\lcorr$ and $\widehat{w}$ is a white noise in Fourier space:
\be
\langle \hat{w}^*(k_1) \hat{w}(k_2) \rangle = 2\boldsymbol{\delta}(k_1-k_2), \qquad
\langle \hat{w}(k_1) \hat{w}(k_2) \rangle = 0;
\label{t5_01b}
\ee
\label{t5_01}
\esube
here $\langle\cdots\rangle$ stand for the ensemble averaging and 
$\boldsymbol{\delta}$, for the delta-function.

As we have announced in the previous section, 
we will be interested in an evolution that
leads to the emergence 
of an intense pulse from an initially disordered state \eqref{t5_01}. 
Due to nonlinear Landau
damping, spectral components of the evolving electric field's envelope
 shift off-center during the evolution,
which causes the forming pulse to move (with non-zero acceleration) 
in the reference frame of
Eq.~\eqref{tn_04}. 
Modeling realistically such a moving pulse would require a very large
spatial computational domain, 
which is out of reach for our computational resources. 
We therefore resorted to a common numerical trick: We impose periodic boundary 
conditions on a finite-length domain, 
thereby modeling repeated passing of a pulse 
(or some disordered field) through this domain. 
For the simulations reported in the main part of this work,
we restricted our consideration to
the nondimensional value $L=40\pi\approx 126$. 
In Appendix B we demonstrate that increasing the length of 
the computational domain does not alter our principal finding,
which is the formation of an intense pulse from an initially 
disordered field due to nonlinear Landau damping (see next section).
Moreover, we explain there why the periodicity of the boundary conditions
does not affect this formation at all, as long as the domain is
sufficiently long.

Thus, in our simulations we have four physical parameters that we varied:
ratio $r$ of the random and constant components in the initial state of the field
\eqref{t5_01}; correlation length $\lcorr$ of the random component there; 
nonlinear Landau damping coefficient $(s/q)$, 
and the cubic nonlinearity coefficient $Q$ in \eqref{tn_04}.

\subsection{Main results}
\label{mr}

\subsubsection{Formation of an intense pulse due to nonlinear Landau damping}
\label{mrs1}

We will now present 
the first of the two key findings
of this study: the formation, due to
nonlinear Landau damping,  of a long-living
intense pulse from a disordered initial field. 
At this point, it is appropriate to remind the reader 
(see, e.g., \citealt{2007Natur.450.1054S}; \citealt{2010PhFl...22c6601F};
\citealt{2015PhLA..379.1821L}; \citealt{15_AgafontsevZakharov}; 
\citealt{18_GelashAgafontsev})
that in
the purely cubic NLSE, intense pulses do routinely emerge from a
disordered state, with this emergence occurring on a time scale 
that is several times faster than $O(1/Q)$. However, they
dissolve about as quickly as they emerge. (Typically, the greater
the pulse's amplitude, the shorter time it ``lives".) 
In contrast to this ``flickering pulse" behavior in the purely
cubic NLSE, in the NLSE with nonlinear Landau damping an intense
pulse forms over a time on the order of $O(1/Q)$ and then propagates
stably over a long time without showing any sign of decay.
We will now present a detailed description of this process.

\begin{figure}
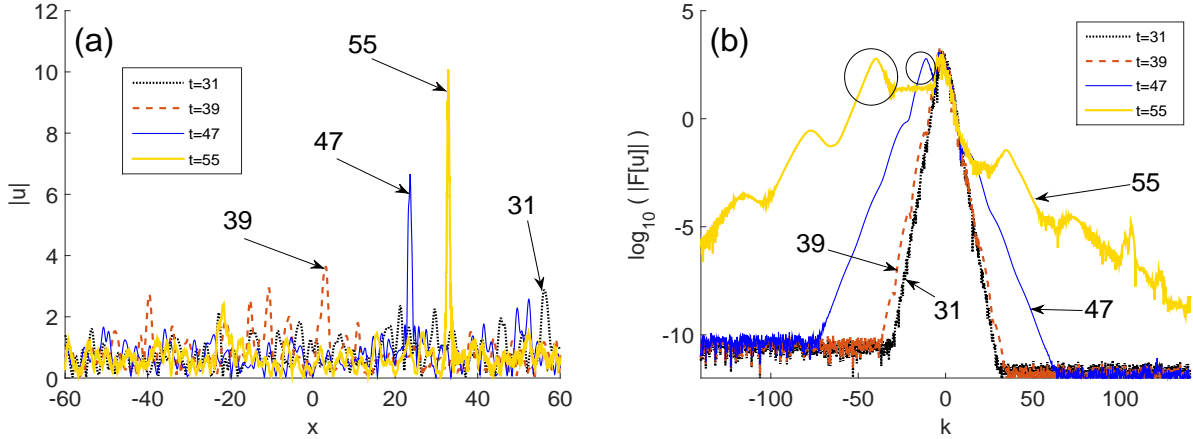

\begin{minipage}{7.5cm}
\hspace*{-0.1cm} 
\includegraphics[height=6cm,width=7.5cm,angle=0]
{figpap_tn1a.eps}
\end{minipage}
\hspace{0.5cm}
\begin{minipage}{7.5cm}
\hspace*{-0.1cm} 
\includegraphics[height=6cm,width=7.5cm,angle=0]
{figpap_tn1b.eps}
\end{minipage}
\caption{
(Color online) \ Field (a) and spectrum (b) of the solution of Eq.~\eqref{tn_04} with the
initial condition \eqref{t5_01} for
$Q=0.25$, $r=1$, $\lcorr=\pi$, and $s/q=0.05$. 
(Only part of the spectral domain is shown for better visibility
of details.) The arrows mark the times at which the solution is plotted. 
The circles in panel (b) mark the ``solitonic" part of the pulse, corresponding
to the intense peak, as discussed in the text. 
}
\label{fig_t1}
\end{figure}

A typical evolution of the field and its Fourier
spectrum, representative of such a process, is illustrated 
in Fig.~\ref{fig_t1}. 
At first, there is a transitional time interval 
(approximately until $t=40$ in the
case shown in Fig.~\ref{fig_t1}), during which the field and its spectrum
 remain statistically similar to the initial ones. 
Some moderately intense
field fluctuations occur during that time, but they keep on
quickly dissolving,
thus mimicking a disordered field evolution in the purely cubic NLSE 
described in the previous paragraph.
Then, a {\em new phenomenon} appears due to nonlinear Landau damping:
Within a short period (somewhere in 
$40 \lesssim t < 45$ in the case of Fig.~{\ref{fig_t1}} (b)), an intense pulse
begins to form and, most importantly, 
{\em no longer dissolves back into a disordered state}.  
As the pulse keeps on becoming taller and narrower, 
its spectrum develops a secondary peak (circled in Fig.~\ref{fig_t1}), 
which begins to shift exceedingly fast away from the original central
wavenumber of the field. 
This shift of the field's spectral components occurs due to 
nonlinear Landau damping,
whereby energy from Fourier harmonics with $k>0$ is transferred to
those with $k<0$ (for $(s/q)>0$ in Eq.~\eqref{tn_04}). 
This stage, where the pulse ``matures", 
takes a relatively short time (approximately corresponding
to $45<t<55$ in the case of Fig.~\ref{fig_t1}), after which 
the growth and narrowing of the pulse
in physical space slow down and then cease.
In the ``mature" stage, the only two effects of 
the nonlinear Landau damping on the pulse evolution are:
the accelerated moving (in the reference frame of Eq.~\eqref{tn_04})
in the physical space and 
the moving of the secondary peak in the Fourier space. 
For the parameters used in this simulation, one is able to reliably
observe this ``mature" stage of the pulse evolution only for a relatively
short time. The reason is that by $t\sim 65$, 
the secondary peak has already moved too close to 
the left edge of the computational spectral domain,
so that spectral components of the solution
near $k\approx -k_{\max}$ have increased considerably above the initial noise level.
To prevent those components from invalidating the numerical solution, we
stopped simulations when the Fourier amplitude of those components would
reach (an arbitrarily chosen) value $10^{-4}$.
A detailed observation and examination of the ``mature" stage 
of the pulse evolution is possible either for a wider computational
spectral domain (see below) or 
for smaller values of $Q$ or $(s/q)$. However, the latter would increase the
time needed for the pulse formation to several hundred units, which 
is beyond the physically relevant range
{\eqref{tn_05b}},  and therefore is not presented here.

The unbounded widening of the solution's spectrum presents an
issue for the validity of that solution not only from the 
numerical, but also from a physical perspective.
Indeed, one of the key assumptions
under which the governing equation {\eqref{NLSwdamp}}
 is valid is that the
characteristic scale of the initial perturbation in the plasma must be
much greater than the Langmuir wavelength. 
After Eqs.~{\eqref{tn_01}}, we assumed that the ratio of these two 
spatial scales, denoted there as $\theta/(2\pi)$, is on 
the order of $10^2\,\ldots\,10^3$.
When the solution's spectrum widens $M$ times, its characteristic spatial scale 
decreases by the same factor. Therefore, the governing model remains valid only
as long as $\theta/(2\pi M) \gg 1$, whence one must require that the spectrum
widening factor be limited by $M<10^2$. If the initial 
(nondimensional) spectral half-width of
the solution is $k\sim 2$, as in Fig.~{\ref{fig_t1}}, then the solution will
remain \emph{physically} valid as long as the separation between the secondary
and primary spectral peaks does not exceed $\sim 100$ units.

A closer examination of the solution's Fourier spectrum reveals
 that while the remnants of the original
spectral peak remain ``noisy" (i.e., jagged in Fig.~\ref{fig_t1}),
the secondary, ``breakaway" peak, corresponding to the intense pulse,
 is smooth and has clearly seen exponentially decaying tails. 
This leads us to hypothesize that the created pulse is a soliton of the perturbed
NLSE \eqref{tn_04}. Creation of a
{\em long-living} intense pulse out
of a disordered initial condition has been reported before 
(\citealt{01StatEquilibGNLS}; \citealt{beta3}) 
for the generalized NLSE. 
However, in those cases, the size of the perturbation, interpreted as the
difference between the given NLSE and the purely cubic one, was not small.
Namely, either the nonlinearity was {\em considerably} different
from cubic (\cite{01StatEquilibGNLS} and references therein) or
a higher-order dispersion term in the cubic NLSE was of order one
(\citealt{beta3}).
Moreover, there are two differences in our observation of 
the pulse's emergence compared to 
such observations in those earlier studies. First, our Eq.~\eqref{tn_04} contains
only a {\em small} perturbation to the NLSE: $(s/q) \lesssim 0.1$.\footnote{
	We observed pulse formation even for values of $(s/q)<0.01$,
	but it occurs over proportionally longer times, which are outside the 
	physically relevant range \eqref{tn_05b}.
	}
 This makes the second 
difference even more surprising: in strongly non-cubic generalized NLSE
considered earlier by \cite{01StatEquilibGNLS}, 
the time that it took an intense pulse to emerge was about 
two to three
orders of magnitude {\em greater} than in the {\em slightly} perturbed NLSE 
\eqref{tn_04}. 
A theoretical explanation of these differences, as well as of the very fact that
a small nonlinear Landau damping causes the emergence of an intense pulse from a disordered
state, remains an open problem. 
At a qualitative level, the emerging pulse appears to harvest energy from the
surrounding field. However, what triggers the event after which the pulse
begins doing so, and why its growth eventually ceases, is unknown.

Thus, to summarize our first key finding: A small nonlinear Landau damping
leads to two qualitative changes of a disordered initial field.
First, unexpectedly, it causes formation of an intense pulse that exists
over a long time (at least as long as the model {\eqref{NLSwdamp}} remains
physically and numerically valid). 
Second, expectedly, it leads to an
(accelerated) shift of that pulse in the spectral domain towards lower
 wavenumbers (for $(s/q)>0$).


\subsubsection{Internal structure of the pulse, and features of radiation}
\label{mrs2}

\begin{figure}
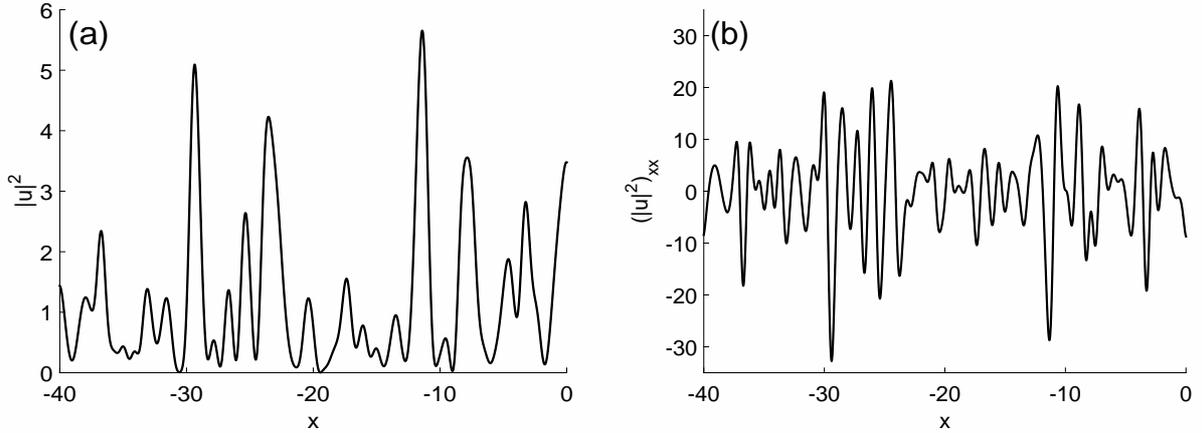

\begin{minipage}{7.5cm}
\hspace*{-0.1cm} 
\includegraphics[height=6cm,width=7.5cm,angle=0]
{fig102901_initial_usq.eps}
\end{minipage}
\hspace{0.5cm}
\begin{minipage}{7.5cm}
\hspace*{-0.1cm} 
\includegraphics[height=6cm,width=7.5cm,angle=0]
{fig102901_initial_usqxx.eps}
\end{minipage}
\caption{
Quantities proportional to the intensities of the electric field (a)
and radiation (b) in the initial state ($t=0$). Simulation parameters are
the same as in Fig.~\ref{fig_t1}, except that the computational spectrum is
three times broader. Only part of the spatial computational domain
is shown for clarity.
}
\label{fig_t2a}
\end{figure}

\begin{figure}
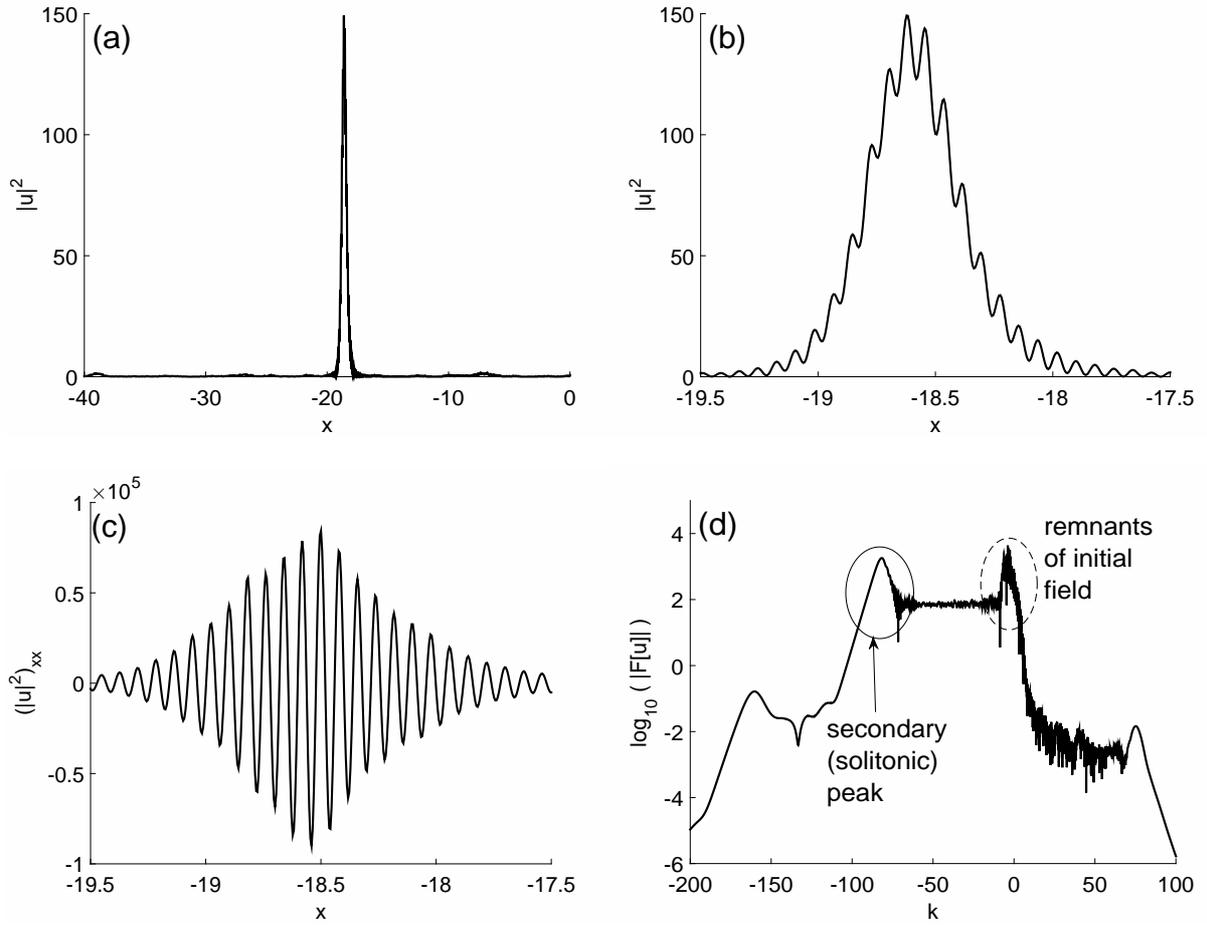

\begin{minipage}{7.5cm}
\hspace*{-0.1cm} 
\includegraphics[height=6cm,width=7.5cm,angle=0]
{fig102901_t65_usq.eps}
\end{minipage}
\hspace{0.5cm}
\begin{minipage}{7.5cm}
\hspace*{-0.1cm} 
\includegraphics[height=6cm,width=7.5cm,angle=0]
{fig102901_t65_usq_closeup.eps}
\end{minipage}

\bigskip

\begin{minipage}{7.5cm}
\hspace*{-0.1cm} 
\includegraphics[height=6cm,width=7.5cm,angle=0]
{fig102901_t65_usqxx_closeup.eps}
\end{minipage}
\hspace{0.5cm}
\begin{minipage}{7.5cm}
\hspace*{-0.1cm} 
\includegraphics[height=6cm,width=7.5cm,angle=0]
{fig102901_t65_spectrum.eps}
\end{minipage}
\caption{
Same simulation as in Fig.~{\ref{fig_t2a}}, but at $t=66$,
after an intense pulse has formed and ``matured". Panel (a) is a 
counterpart of that in Fig.~{\ref{fig_t2a}}. 
Note that most of the computational domain is occupied by the field whose 
intensity is some two orders of magnitude smaller than that of the 
main pulse.
Panel (b) is a 
close-up of (a) focusing on the vicinity of the pulse. Panel (c)
is a counterpart of Fig.~{\ref{fig_t2a}}(b). Panel (d) shows 
part of the Fourier spectrum of the numerical solution.
}
\label{fig_t2b}
\end{figure}

Our second key finding concerns the \emph{wavelength} of the radiation
emitted by the plasma where an intense pulse has formed due to the mechanism
described above. The intense pulse creates a ponderomotive force which prevents
the charge bunch from collapsing. The ponderomotive force is 
$\propto \nabla |E_{\|}|^2$,
which, when used in conjunction with the Poisson equation and restricted to the
one-dimensional motion along magnetic field lines (see section 1), 
gives the charge density across the pulse to be proportional 
to $\partial^2 |E_{\|}|^2/\partial x^2$ (see 
MGP00). The charge density, in turn, is a coherent structure bounded by the width of
the intense pulse, which moves along curved magnetic field lines to produce
coherent curvature radiation. To illustrate that finding, in 
Figs.~{\ref{fig_t2a}} and {\ref{fig_t2b}}
we compare the quantity $\partial^2 |u|^2/\partial x^2$ (recall $u$ is the 
nondimensional electric field given by Eq.\eqref{tn_02}),
 in the initial state (Fig.~{\ref{fig_t2a}}) 
and in a state where a pulse has formed (Fig.~{\ref{fig_t2b}}). 
The simulation was run for the same parameters as in Fig.~{\ref{fig_t1}},
except that, in order to resolve a high-wavenumber ripple in 
Fig.~{\ref{fig_t2b}}, we used a three-time wider spectral window
and a correspondingly smaller time step (see Appendix A). 
Since the time of formation of an intense pulse is sensitive to the
initial condition (and hence the computational spectrum), 
in Figs.~{\ref{fig_t2a}} and ~{\ref{fig_t2b}} it is
different from that in Fig.~{\ref{fig_t1}}.

A comparison of Fig.~{\ref{fig_t2a}}(b) and Fig.~{\ref{fig_t2b}}(c)
shows a more than three-order of magnitude increase of the 
radiation's intensity.
It is important to note that this increase occurs due to two
separate reasons: first, formation of a pulse whose amplitude, i.e. $\max |u|$,
 is an order of magnitude greater than that of the initial field
(compare Fig.~{\ref{fig_t2a}}(a) to Fig.~{\ref{fig_t2b}}(a)), and 
second, existence of a highly oscillatory ripple ``on top" of the pulse
(see Fig.~{\ref{fig_t2b}}(b)). The spatial period of this ripple is
about an order of magnitude smaller than unity, which adds approximately
two orders of magnitude to the size of the second derivative of $|u|^2$
(see Fig.~\ref{fig_t2b}(c)).

The wavelength of the ripple $\delta x_{\rm ripple}$ ``on top" of the intense pulse
is explained by the spectrum of the 
field: see Fig.~{\ref{fig_t2b}}(d). Specifically, the ``bulk" of the
pulse corresponds to the ``solitonic" spectral peak located at $k\approx -80$
in that figure. On the other hand, ``remnants" of the initial field have
$k\approx 0$. Thus, the intensity of the superposition of these two parts
of the field, i.e.
\be
|u|^2 \approx \big| A_1 e^{-i80x} + A_2 e^{i0x} \big|^2 =
(|A_1|^2+|A_2|^2) + 2|A_1||A_2|\cos(80x+\phi),
\label{e2_01}
\ee
has the approximate wavelength of the ripple as, 
$\delta x_{\rm ripple}= (2\pi)/80\approx 0.07$.
(In {\eqref{e2_01}}, $A_1, A_2$, and $\phi$ are some constants.)
This wavelength is the smallest scale of a coherent structure inside the 
intense solitonic pulse and is clearly visible in Fig.~\ref{fig_t2b}(b) and \ref{fig_t2b}(c).
In section 7 we will demonstrate that while this structure is ``short-lived"
compared to the pulse itself, it still can be a source of coherent emission.

Thus, we have found two types of structures that emerge from an initially
disordered state of electric field in the pulsar plasma: \ 
(i) an intense pulse of the envelope of Langmuir waves and \ (ii) a ``ripple"
on top of this pulse. We will conclude this subsection with an estimation of 
their characteristic spatial scales and the corresponding frequencies that they
can emit. We will begin with the intense pulse and use the parameters
reported for, and shown in, Fig.~\ref{fig_t2b}. 
In dimensional units, the width of the pulse, 
$\delta \chi_{\rm pulse,\,\circ}\,$, in the OFR can be estimated as follows.
First, since the nondimensional time that it takes the intense pulse to form is
$t\approx 60 \sim 100$, then \eqref{tn_05b} implies that $\theta\sim10^3$.
Using this value and $\omega_p \sim 4\times 10^{9}$ s$^{-1}$
 in Eqs.~{\eqref{tn_01}} 
and a value $\gamma_p \sim 10^2$  to convert between the OFR and MFR variables, one obtains:
\bsube
\be
\delta \chi_{\rm pulse,\,\circ} = \delta \chi/\gamma_p = 
l \theta \delta x_{\rm pulse}/\gamma_p \sim
3\times 10^8 ~~{\rm m\cdot s}^{-1} / (4\times 10^9 ~~{\rm s}^{-1} )
 \cdot 10^3 \cdot 1 / 10^2 \sim 70~~{\rm cm}.
\label{e2_02a}
\ee
This corresponds to the radiation's frequency of about 
\ $c/\delta \chi_{\rm pulse,\,\circ} \sim 400$ MHz
and thus falls in the mid-range of the spectrum of observed 
pulsar coherent radio emission, which extends from several tens of
MHz to a few GHz. 
Let us note that this estimate can go down by a factor of two or so
if one allows for the possibility that the radiation is emitted from
altitudes $R_{em}$ that are lower than the value of $500$ km assumed in \eqref{Rrad};
see the paragraph before Eq.~\eqref{tn_02}.
Similarly, since the nondimensional wavelength $\delta x_{\rm ripple}$ of the
ripple on top of the pulse is seen to be about an order of magnitude smaller, then
\be
\delta \chi_{\rm ripple,\,\circ} \sim 7~~{\rm cm},
\label{e2_02b}
\ee
\label{e2_02}
\esube
and the corresponding frequency is about 4 GHz. 
In the next subsection we will show that both frequencies following from estimates
\eqref{e2_02} may go down by about an order of magnitude if one assumes a higher 
value of the nonlinearity coefficient. 
In section 7 we will further discuss the relevance of our numerical observations
to the problem of coherent emission by change bunches in plasma. 

\smallskip

Having presented our main findings, we now describe how the formation
of an intense pulse is affected by the physical parameters of the
governing equation {\eqref{tn_04}}.

\subsection{Dependence of pulse formation on $Q$}
\label{depq}

Predictably, as one increases $Q$ without changing other parameters in \eqref{tn_04},
the field evolution due to nonlinear terms (both purely cubic and 
Landau-damping ones) occurs faster, and the time required for the pulse formation
decreases. This is illustrated in Table \ref{tab_t1},  where 
we list the times $t_{4\times}$ that it takes the pulse's amplitude,
$\max_x |u|$, to exceed an arbitrarily set threshold of 4, 
which is four times the average amplitude of the
initial state \eqref{t5_01}.
These times are seen to decrease with the increase of $Q$, as expected.
At the same time, we found that the spatial scale of the intense pulse
is not significantly affected by $Q$. This fact will play a role in the
forthcoming estimate of the dimensional wavelength and frequency of the
coherent radio emission that such a pulse can generate.

\begin{table}
\begin{center}
\begin{tabular}{|l||*{13}{c|}}\hline
\backslashbox{$(s/q)$}{\vspace*{-0.2cm}\hspace*{-0.6cm}$Q$}
& 0.3 & 0.4 & 0.5 & 0.6 & 0.7 & 0.8 & 0.9 & 1.0 & 1.1 & 1.2 & 1.3 & 1.4 & 1.5 
 \\\hline\hline
\quad 0.05 & 
40 & 30 & 20 & 12 & 6.1 & 5.8 & 4.0 & 3.0 & 3.4 & 2.9 & 2.0 & 1.8 & 1.5 \\\hline
\quad 0.10 &
21 & 11 & 7.5 & 5.5 & 4.4 & 3.5 & 3.0 & 3.5 & 2.8 & 2.1 & 2.0 & 1.6 & 1.5  \\\hline
\end{tabular}
\end{center}
\caption{Evolution times $t_{4\times}$ that it takes the pulse's amplitude
to exceed four times the average amplitude of the initial state. 
The initial condition \eqref{t5_01} has parameters $r=0.9$ and $l_{\rm corr}=\pi$.
All simulations use the same seed of the random number generator, leading to the
same initial pulse profile. 
Since details of the pulse evolution depend on the (randomly chosen) initial
profile, the times $t_{4\times}$ are listed only to two significant figures, 
which suffices to illustrate the general trend.
 }
\label{tab_t1}
\end{table}

Namely, we can accept that the distance $\Delta R$ that it takes for the instability in
the secondary plasma cloud to lead to formation of an intense pulse is
given by \eqref{rinl}; then the {\em dimensional} time of pulse formation
continues to be given by \eqref{tn_05a}. In such a case, the decrease in the
{\em nondimensional} time, seen in Table \ref{tab_t1} as $Q$ increases, 
implies that $\theta$ takes on values from the upper part of its range, e.g.,
$\theta\sim 10^4$: see \eqref{tn_05b}. Now, for a greater $\theta$, a given
nondimensional spatial scale corresponds to a greater dimensional scale:
see \eqref{tn_01}. Thus, as $Q$ increases, the dimensional scale of 
both the solitonic pulse and the fine
structure on top of it can go up by an order of magnitude compared
to \eqref{e2_02}.
Consequently, the frequencies of the coherent emission can be found
in the range from several tens to several hundreds of MHz for the 
pulse and the ``ripple", respectively.

We should be careful to note that this is only one interpretation of the observed
decrease of $t_{4\times}$ with $Q$; other interpretations may be possible.
For example, one can assume that the increase of $Q$ implies that the Langmuir
turbulence in the plasma cloud is so strong that the formation of an intense pulse
occurs not over 500 km, as in \eqref{rinl}, but much sooner, say, over 100 km.
In this case, {\em both} $\theta$ in \eqref{tn_05b} and $R_{50}$ in \eqref{omegap}
would change compared to their values in \eqref{e2_02}. However, a more detailed 
analysis of such a possibility is outside the scope of this study.

Coming back to Table \ref{tab_t1}, we note that for the two different values
of the nonlinear Landau damping coefficient $(s/q)$, 
the most pronounced decrease of  $t_{4\times}$ occurs for a higher range
of $Q$ values for the smaller $(s/q)$: for $Q\in(0.4,\, 1.0)$ for $(s/q)=0.05$,
and for $Q\in(0.3,\, 0.5)$ for $(s/q)=0.10$. For both values of $(s/q)$,
the decrease of $t_{4\times}$ significantly slows down for $Q$ values above
those respective ranges.

In the next subsection we will discuss other changes in the pulse evolution
that occur with changing the nonlinear Landau damping coefficient $(s/q)$.

\subsection{Dependence of pulse formation on $(s/q)$}

\begin{figure}
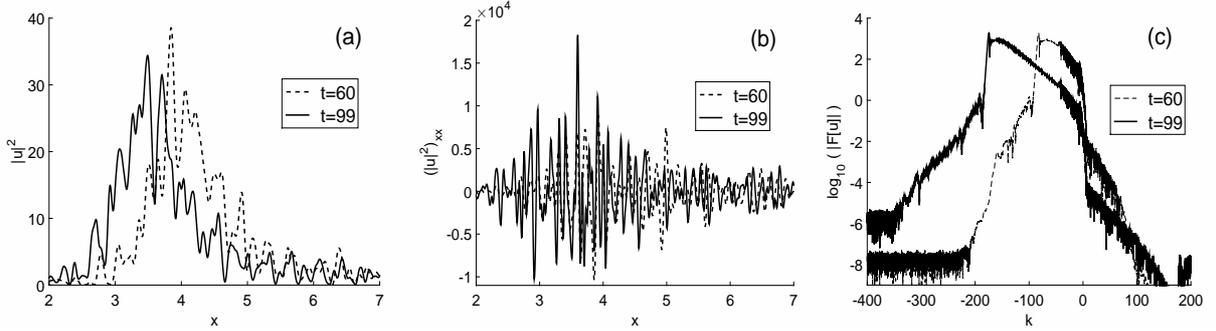
 
\begin{minipage}{5cm}
\hspace*{-0.2cm} 
\includegraphics[height=4.4cm,width=5cm,angle=0]
{fig103001_t60and99_usq_closeup.eps}
\end{minipage}
\hspace{0.2cm}
\begin{minipage}{5cm}
\hspace*{-0.1cm} 
\includegraphics[height=4.4cm,width=5cm,angle=0]
{fig103001_t60and99_usqxx_closeup.eps}
\end{minipage}
\hspace{0.2cm}
\begin{minipage}{5cm}
\hspace*{-0.1cm} 
\includegraphics[height=4.4cm,width=5cm,angle=0]
{fig103001_t60and99_spectrum.eps}
\end{minipage}
\caption{Same simulation parameters as for
Fig.~\ref{fig_t2b}, but for a larger value of 
nonlinear Landau damping $(s/q)=0.15$.
Solution at two times are shown to illustrate slow evolution of the 
field's spectrum for larger values of $(s/q)$. The two times are chosen
so that the locations of the intense pulse would nearly coincide, to
facilitate visual comparison of the two pulses. 
Only part of spatial and spectral domains is shown for better visibility.
}
\label{fig_t3}
\end{figure}

The expected effect of varying the nonlinear Landau damping coefficient
 is that the formation time
of an intense pulse decreases as $(s/q)$ increases. However,
and perhaps less expectedly, 
an increase of nonlinear Landau damping beyond a certain point 
begins to have less effect on the speed of the pulse formation.
This can be seen from the data reported for larger values of $(s/q)$ 
in Table \ref{tab_t2}.
Moreover, at least four other changes occur with the increase of 
nonlinear Landau damping.
First, the shape of the pulse becomes visibly asymmetric, with a ``tail" 
forming behind the pulse; see Fig.~{\ref{fig_t3}}(a). 
Second, this change in the shape is accompanied by a decrease 
of the amplitude of the ``matured" pulse; compare Fig.~{\ref{fig_t3}}(a) to
Figs.~{\ref{fig_t2b}}(a,b). 
Third, the intensity of the radiation emitted by the pulse decreases,
whereas its wavelength increases:
compare Fig.~{\ref{fig_t3}}(b) to Fig.~{\ref{fig_t2b}}(c)
and note that respective horizontal and vertical scales are different. 
These effects are manifested in the Fourier space
as follows: \ (i) \ the spectral peak corresponding to the intense pulse is
now much less prominent over a spectral ``plateau" that is observed
immediately on its right side, and \ (ii) ``remnants" of the initial field
with spectral components near $k=0$ have been considerably reduced for the
larger value of $(s/q)$. Thus, the spectrum of the field is considerably
narrower for the larger values of nonlinear Landau damping.
Finally, and perhaps unexpectedly, the shift of the spectrum occurs 
much slower for larger values 
of $(s/q)$. Specifically, the pulse shown in Fig.~{\ref{fig_t3}} for
$(s/q)=0.15$ forms around $t=25$, whereas that shown in
Fig.~{\ref{fig_t2b}} for $(s/q)=0.05$ forms around $t=50$. The spectrum
of the latter pulse approaches the left edge of the computational domain
(i.e., $-k_{\max}\approx -600$ in this case) already for $t=75$, by which
time the numerical solution becomes invalid (see the beginning of section 
{\ref{numsim}}).
On the contrary, the spectrum of the pulse for $(s/q)=0.15$ is seen
not to reach even one half of the computational spectral window by
$t=100$. 

\begin{table}
\begin{center}
\begin{tabular}{|l||*{9}{c|}}\hline
\backslashbox{$(Q;\;l_{\rm corr})$}{\vspace*{-0.2cm}\hspace*{-0.6cm}$(s/q)$}
& 0.01 & 0.02 & 0.04 & 0.06 & 0.08 & 0.10 & 0.13 & 0.16 & 0.20 
 \\\hline\hline
$(1;\; \pi)$ & 
14 & 8.0 & 4.5 & 4.1 & 3.5 & 3.4 & 3.1 & 3.0 & 2.6  \\\hline
$(1;\; 1)$ &
$>100$ & 65 & 32 & 27 & 16 & 12 & 12 & 9.2 & 9.4   \\\hline
$(2;\; 1)$ &
$30$ & 14 & 11 & 7.0 & 6.1 & 4.6 & 3.5 & 2.5 & 4.4   \\\hline
\end{tabular}
\end{center}
\caption{Dependence of the 
times $t_{4\times}$ on the nonlinear Landau damping coefficient. 
The initial condition for each simulation is the same and has $r=0.9$.
 }
\label{tab_t2}
\end{table}

For completeness,
we also verified that as $(s/q)$ increases to become of order one, a
pulse no longer forms. Instead, a ``step" with an oscillatory ``tail"
is formed. The spectrum of this solution is approximately flat at the top,
with the top's width increasing with time.

\subsection{Dependence of pulse formation on $\lcorr$}

The effect of the spatial scale of the initial condition
 on the field evolution is predictable, at least withing some range.
Namely, as $\lcorr$ decreases, the effect of the dispersive term $u_{xx}$ in
Eq.~\eqref{tn_04} increases compared to that of the nonlinear term.
Thus, having a smaller $\lcorr$ in the initial condition is, essentially,
similar to having a smaller $Q$: it delays pulse formation. 
This is confirmed by comparing the first and second lines in Table \ref{tab_t2},
which show the effect of decreasing $l_{\rm corr}$.
In comparison, the third and second lines of the same Table show that
a similar effect of pulse formation delay occurs when $Q$ is decreased,
as has already been noted in section 6.2.

\subsection{Dependence of pulse formation on $r$}

\begin{figure}
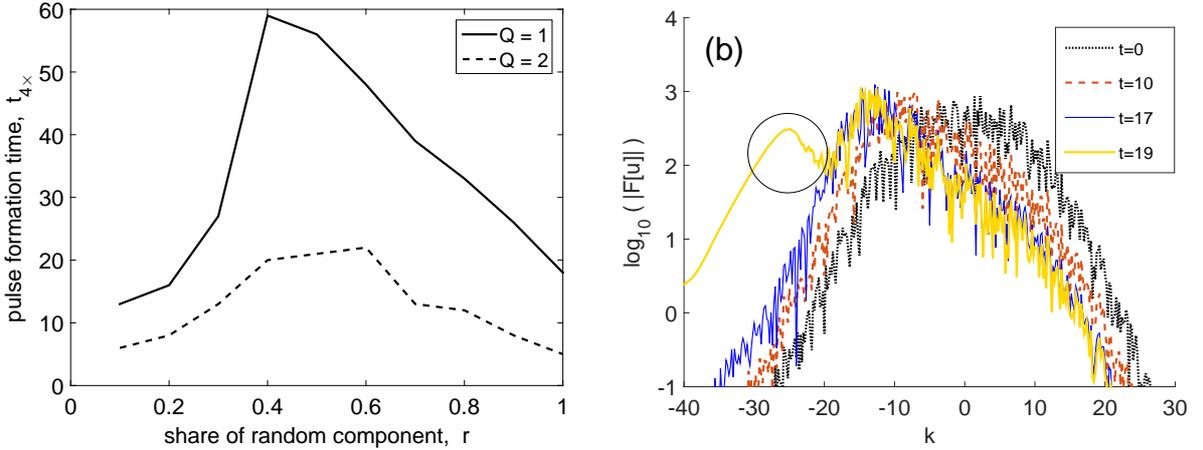

\begin{minipage}{7.5cm}
\hspace*{-0.1cm} 
\includegraphics[height=6cm,width=7.5cm,angle=0]
{figpap_tn4a.eps}
\end{minipage}
\hspace{0.5cm}
\begin{minipage}{7.5cm}
\hspace*{-0.1cm} 
\includegraphics[height=6cm,width=7.5cm,angle=0]
{figpap_tn4b.eps}
\end{minipage}
\caption{(a) \ Typical dependence of the pulse formation time on the share of the random
component in the initial field. Parameters are: $l_{\rm corr}=1$ and $(s/q)=0.05$.
To generate the random part of the initial condition, we used 
the same seed of the random number generator for all $r$. 
\ (b) \ (Color online) \ Fourier spectra illustrating the evolution
of an initial random state ($r=1.0$) into an intense pulse, as explained in the text;
$Q=1$.  
Only part of the spectral domain is shown for better visibility of details.
}
\label{fig_t4}
\end{figure}

Finally, we investigated how the ``degree of randomness"
of the initial state affects the pulse formation.
Initially, we expected that the formation times would increase
with the share of the random component in the initial state. 
However, in numerical experiments we observed that while those times indeed
initially increase with $r$, they reach a maximum around $r=0.5$ and then begin to 
decrease. Representative results are shown in Fig.~\ref{fig_t4}(a). 
We observed qualitatively the same
results for several other values of parameters $Q$ and $(s/q)$
than reported in Fig.~\ref{fig_t4}, 
as well as for a different initial random state (controlled by the 
seed of the random number generator in the numerical code). 

In Fig.~\ref{fig_t4}(b) we show different stages of the field evolution, which were
found to be similar for all $r\gtrsim 0.4$. Initially,
nonlinear Landau damping leads mostly to a (rather slow) shift
of the field's Fourier spectrum, with only minor changes of the spectrum's shape;
compare the curves for $t=0$ and $t=10$. 
 In physical space, the field appears disordered
during that stage. 
Then, the spectrum begins to become noticeably narrower and asymmetric;
it is shown at $t=17$, which is shortly before the formation of an intense pulse.
(Incidentally, the spectral narrowing corresponds to the
increase of the correlation length of the field, and, according to the 
previous subsection, this facilitates the pulse formation.)
Finally, an intense pulse forms within a relatively short time interval;
\footnote{
  In other simulations we observed that the duration of this time interval 
  scales inversely proportional to $Q$, but seems to be rather insensitive
  to $(s/q)$ for sufficiently small values of that parameter.
	}
see the curve for $t=19$ in Fig.~\ref{fig_t4}(b).
 In the spectrum, the formation of a pulse 
corresponds to the emergence of a secondary peak around $k \approx -30$,
marked with a circle;
compare it with Figs.~\ref{fig_t1}(b) 
and {\ref{fig_t2b}}(d).
Once the pulse is formed, its height and width remain almost
unchanged, until the numerical solution loses its validity due to 
a significant part of the spectrum shifting near the left edge of the spectral
window (see section 6.1.1).


\section{Summary and  Discussion}
\label{sec6}

We have addressed the open problem of explaining a mechanism of coherent
curvature radio emission by the electron--positron plasma in 
pulsar magnetosphere.
As the mathematical model of this phenomenon we considered the generalized
nonlinear Schr\"odinger equation (NLSE) proposed by 
\citet{2000ApJ...544.1081M} (MGP00), which includes effects of group velocity
dispersion, nonlinearity of electric susceptibility, 
and resonant interaction between
Langmuir waves and plasma particles (nonlinear Landau damping). 
In the absence of nonlinear Landau damping, the purely cubic 
NLSE can, in principle, support
solitons, which in the plasma would be manifested as 
charge bunches that propagate
stably and therefore are capable of emitting coherent radiation. 
However, formation
of solitons in the purely cubic NLSE requires that initially, 
the Langmuir wave have the
envelope that is either localized or consists of several 
well-separated localized ``bumps".
It is only then that the emerging 
charge solitons can maintain their shape for a
sufficiently long time to radiate coherently. There is no reason to expect that
such a special initial condition of Langmuir waves would exist in a disordered 
pulsar plasma. Then, it is known 
(see section 6.1.1) 
that evolution of a {\em disordered initial state}
in the purely cubic
 NLSE leads to an ensemble of strongly interacting pulses, which constantly
appear, disappear, and change their shape due to the interaction. 
Such a disordered,
in both time and space, ensemble of pulses cannot be expected to emit coherently.

Motivated by this 
inability
of the purely cubic NLSE to identify a candidate mechanism of
coherent emission, we numerically solved the NLSE {\em with the nonlinear Landau
damping} term, as derived by MGP00. 
We found that for a range of realistic values of pulsar
parameters, the presence of nonlinear Landau damping leads to 
the formation of an intense,
soliton-like pulse out of an initially disordered Langmuir wave. Such a stable
pulse can emit coherently and thus is a reasonable candidate as a source of
coherent radio emission. 
However, an analytical explanation of this emergence of a long-living intense pulse 
remains an open problem.

Let us point out a key difference between this result and the results of earlier
studies (\citealt{78_NLS_SolPertNonlinLD};
\citealt{89_DNLS_MHD_NonlinLD}; \citealt{09_NLS_Dusty_NonlinLD}; 
\citealt{15_NLS_CollisionlessElectronPositron_NonlinLD}; 
\citealt{17_NLS_Degenerate_NonlinLD}; \citet{18_NLS_RelatElectronIon_NonlinLD}), 
which considered the effect of nonlinear Landau damping on an {\em isolated} soliton. 
Those earlier studies found that such a soliton will experience decay and a
frequency shift of the carries. Both of these phenomena are consequences of the
fact that the nonlocal term in the NLSE \eqref{NLSwdamp} describes 
energy transfer from one side of the pulse spectrum to the other.
In contrast, in our simulations, a pulse emerges from an initially disordered
state and, during its ``maturation" stage, appears to absorbs energy from
the surrounding field.

Two important notes  about this pulse formation are in order. First, 
the nonlinear Landau damping coefficient has to fall in 
a certain range (namely, the lower part
of \eqref{tn_06}). If it is too high, then the intensity of the
 emerging pulse is lower,
or a stable pulse may even not form at all; see section 6.3. 
On the other hand, if the
nonlinear Landau damping is too low, the pulse may not have the 
time to form during the stage
when the charge density in the plasma cloud is sufficiently 
large to produce strong radiation; see section 5.

Second, the intense pulse, formed for appropriate values of 
the nonlinear Landau damping coefficient, 
has an internal structure whose spatial scale 
can be about an order of magnitude smaller than the 
spatial extent of the pulse itself;
see Fig.~\ref{fig_t2b}(b). 
In this work we did not undertake an actual {\em calculation} 
of the coherent emission
by such stable pulses, containing a large number of charged particles; 
this clearly requires a separate study.
Without such a calculation, 
one cannot tell to what extent each of these structures:
the ``bulk" solitonic pulse itself and the finer ``ripple" on top of it, 
contribute to the coherent emission. 
It appears intuitively plausible that frequencies in the lower end of the 
observed spectrum (tens to hundreds of MHz) are generated by the
pulse as a whole, while frequencies from the higher end (up to several GHz)
are generated by the ``ripple". 
This is because the spatial scale of the pulse is about an order 
of magnitude greater than
that of the ``ripple"; see sections 6.1.2 and 6.2.
However, a {\em calculation} of the spectrum
emitted by such a {\em two-scale} structure of charges remains an open problem.

\begin{figure}
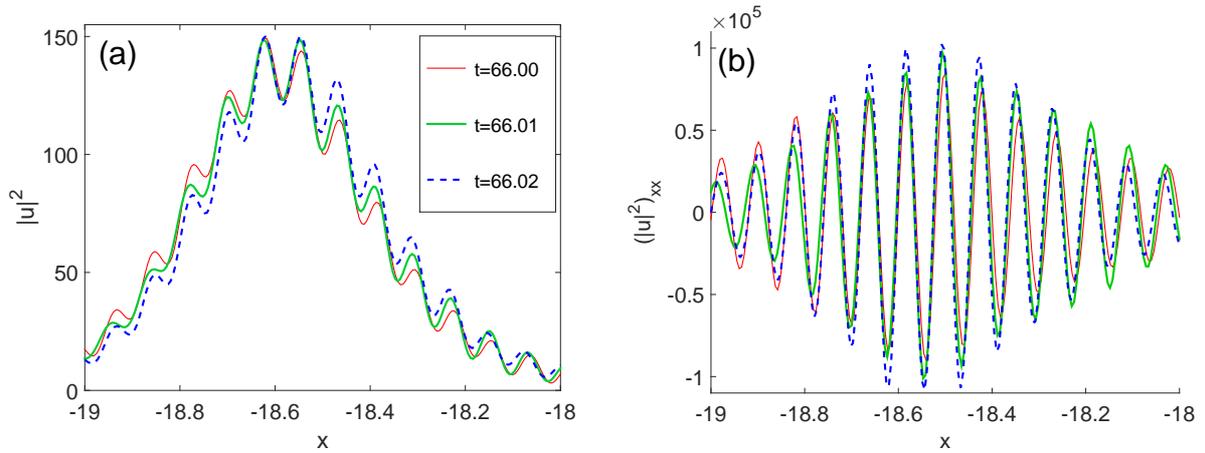

\begin{minipage}{7.5cm}
\hspace*{-0.1cm} 
\includegraphics[height=6cm,width=7.5cm,angle=0]
{figpap_tn5a.eps}
\end{minipage}
\hspace{0.5cm}
\begin{minipage}{7.5cm}
\hspace*{-0.1cm} 
\includegraphics[height=6cm,width=7.5cm,angle=0]
{figpap_tn5b.eps}
\end{minipage}
\caption{(Color online) \  
Same quantities as in Fig.~\ref{fig_t2b}(b),(c), respectively, but for 
$t=66.00,\,66.01,\,66.02$. Line colors and styles are as shown in the
legend in panel (a). The $x$-window is smaller than in Fig.~\ref{fig_t2b}(b),(c)
in order to make the details appear more clearly. 
In both panels, 
the centers of the pulses at the different times are manually superimposed
in order to clearly show the changes of the profile. (If a pulse moves without
changing its shape, its does not affect its ability to emit coherently.) 
We also observed that at $t=66.10$ the profile of the ``ripple" has changed
completely relatively to that at $t=66.00$; the corresponding curves are not
shown in order not to clutter the picture. 
}
\label{fig_t5}
\end{figure}

Let us now demonstrate that while the ``ripple" on top of the solitonic pulse
keeps changing its shape on a time scale that is small compared to the 
time scale where such a pulse exists, 
those changes are still ``slow enough" to allow the ``ripple" 
to emit coherently in the 
range of frequencies estimated in section 6.1.2 (several GHz), and even at 
lower frequencies. 
To that end, note that in order for the ``ripple" to be 
a source of coherent radiation, 
it must exist long enough to guarantee condition \eqref{contr2a}.
Namely, the time ${\mathcal T}_b$  over which the shape of 
this ``ripple" remains 
mostly unchanged must be much greater than the period ${\mathcal T}_c$
of the coherent radio emission. Let us demonstrate, 
using the illustrating example
of Fig.~\ref{fig_t2b}, that this is indeed the case. 
In Fig.~\ref{fig_t5} we show that the profiles of both the electric field's
intensity $|u|^2$ and the ponderomotive force $|u|^2_{xx}$  
 are mostly preserved over $t\approx 0.01$. 
Now, if $t\sim 100$ corresponds
to $500$ km (see section 5), then ${\mathcal T}_b\sim 0.01$ 
corresponds to about $50$ m.
Then, condition \eqref{contr2a} implies that the lower 
limit of frequencies $\omega_c$
is about $c/(50\;{\rm m})\sim 10$ MHz. This is consistent (within a two-order of
magnitude margin)
with the value of several GHz mentioned after estimate \eqref{e2_02b}.

Note that since the solitonic pulse itself stably propagates
over $t>O(10)$ nondimensional units, there is, for practical purposes, 
no lower limit
from condition \eqref{contr2a} 
on the frequencies that it, as a whole, can emit coherently.

Finally, let us note that since we had to use periodic boundary conditions in our
numerical simulations (see the preamble to section 6), 
we always observed that only
one intense pulse forms as a result of many collisions with smaller pulses. 
In an actual plasma cloud, where the pulse passes 
through it only once rather than
repeatedly, many well-separated and long-living solitonic pulses may form.
Then, taking into account emission by this ensemble of stable charge bunches,
as opposed to by a single charge bunch, is yet another open problem.

\section*{Acknowledgments}
We thank the anonymous reviewer for constructive comments. \ 
The work of T.I.L. was supported in part by the NSF grant  DMS-1217006.
D.M. acknowledges funding from the grant  ``Indo-French Centre for the Promotion 
of Advanced Research – CEFIPRA".
D.M. and G.M. would like to thank the Physics 
Department of the University of Vermont,
where this project was started when D.M. held a visiting faculty position there
working in collaboration with Prof J.~Rankin. 
D.M. also thanks Prof. V.N.~Kotov for useful discussions and encouragement during
his work on this project.

\bibliographystyle{apj}
\bibliography{landau_damping_mnras_resubmJuly2018} 


\section*{Appendix A: Numerical method for solving Eq.~(20)}  

We will present this method for
equations of the form
\be
iu_t+\calL u + \calN =0,
\label{appnum_01}
\ee
where $\calL$ is a linear operator with spatially constant coefficients
and $\calN$ includes all other terms. The generalized NLS \eqref{tn_04}
is a special case of \eqref{appnum_01}, with $\calL=\partial^2/\partial x^2$ and
$\calN$ being the entire nonlinear term.
Below we will label the Fourier transform of any quantity
with an over-hat:
$$
\uhat(k,t) = \frac1{\sqrt{2\pi}}\int_{-\infty}^{\infty}
u(x,t) \,e^{-ikx} dx, \qquad
u(x,t)  = \frac1{\sqrt{2\pi}}\int_{-\infty}^{\infty} 
\uhat(k,t) \,e^{ikx} dx.
$$
Similarly, $\calLhat\equiv \calLhat(k)$ and $\calNhat\equiv \calNhat(k)$
will denote the Fourier symbols of operators $\calL$ and $\calN$, respectively. 
For example, in \eqref{tn_04}, $\calLhat=-k^2$.
Taking the Fourier tranform of
\eqref{appnum_01} yields
\be
i\uhat_t+\calLhat\uhat + \calNhat = 0.
\label{appnum_02}
\ee
Solving this from $t_1$ to $t_2$ as a linear inhomogeneous equation
yields:
\be
e^{-i\calLhat t_2}\,\uhat(t_2) - e^{-i\calLhat t_1}\,\uhat(t_1) = 
\int_{t_1}^{t_2} e^{-i\calLhat t'}\,i \calNhat(t')\, dt',
\label{appnum_03}
\ee
where we have suppressed the obvious $k$-dependence of variables. 
Thus, the linear term in \eqref{appnum_01} is
accounted for exactly by \eqref{appnum_03}.

In \cite{2016JSci..72.14L} it was proposed to use the leap-frog scheme
to discretize the integral on the r.h.s. of Eq.~\eqref{appnum_03}
and thereby turn that equation into a numerical method.
The leap-frog scheme is well-known to quasi-preserve\footnote{
   i.e., possibly allow to fluctuate about the initial value,
	 but not shift systematically
	 }
the $L_2$-norm of the numerical solution. 
Choosing it is, therefore, appropriate for the wide
class of equations \eqref{appnum_01} where this norm
is conserved; the generalized NLS \eqref{tn_04} with nonlinear Landau damping
belongs to that class. 
It should also be noted that the leap-frog scheme is explicit
and hence easy to implement.

Within the leap-frog scheme, two versions of the discretization of 
the integral in \eqref{appnum_03} are still possible. As discussed in
\cite{2016JSci..72.14L}, one of them significantly distorts the solution's
spectrum at the edges of the spectral computational window,
while the other does not. For our purposes of modeling nonlinear Landau damping,
which constitutes a transfer of energy from higher- to lower-$k$
Fourier harmonics (for $(s/q)>0$), it is essential to have the spectrum
undistorted at the edges. Therefore, in this study we used the latter
method, which was called IF-LF (integrating factor--leap-frog) in
\cite{2016JSci..72.14L}.\footnote{
	Let us note, in passing, that the other method, which distorts the
	spectrum at the edges, has advantages over the IF-LF in numerical stability.
	}
The form of the IF-LF method is \cite{2016JSci..72.14L}:
\bsube
\be
e^{-i\calLhat t_{n+1}}\,\uhat(t_{n+1}) - e^{-i\calLhat t_{n-1}}\,\uhat(t_{n-1})
  = 2\dt\, e^{-i\calLhat t_{n}} 
	\, i\calNhat(t_n)\,,
\label{appnum_04a}
\ee
where $\dt$ is the time discretization step. 
For the generalized NLS \eqref{tn_04} this simplifies to:
\be
e^{ik^2\dt }\,\uhat(t_{n+1}) - e^{-ik^2\dt}\,\uhat(t_{n-1})
  = 2i\dt\,\calNhat(t_n)\,.
\label{appnum_04b}
\ee
\label{appnum_04}
\esube
This method has accuracy $O(\dt^2)$ in time; the discrete Fourier transform
yields an exponential accuracy in space, provided that the solution with all
its derivatives is continuous.

An extra step is now required to turn method \eqref{appnum_04} into a
useful tool. Namely, if implemented just as above, the method will 
become numerically unstable over a time $t_{\rm inst}$,
which is on the order of a hundred time units for $Q=O(1)$. This numerical instability 
occurs for low-$k$ Fourier harmonics and is 
caused by a so-called ``parasitic" solution, which is well-known to 
be engendered by the leap-frog scheme. As discussed in \cite{2016JSci..72.14L},
for $Q<0$ this instability is essentially a linear, modulational-type
instability. On the other hand, for $Q>0$, the instability is nonlinear
and, to our knowledge, was first analyzed in 
\citet{83_Newell, 86_JCP}. 
A method to suppress the instability caused by the ``parasitic" 
solution was demonstrated and extensively tested in \cite{2016JSci..72.14L}.
It consists of averaging the solution every $t_{\rm stab}$ time units,
with $\dt \ll t_{\rm stab} \ll t_{\rm instab}$, in a way that 
distorts the solution by a negligible amount. 
In this study, we used $t_{\rm stab}=1$.
We will now describe the aforementioned averaging procedure which
we used to stabilize the numerical solution.

Denote $\vhat_n = \uhat_n e^{ik^2\,t_n} \equiv \uhat_n e^{ik^2\,n\dt}$.
Note that \eqref{appnum_04b} is then rewritten as:
\be
\vhat_{n+1} - \vhat_{n-1} = 
2i\dt\,e^{ik^2\,t_n}\,\calNhat(t_n)\,,
\label{appnum_05}
\ee
where $\calNhat(t_n)$ depends on $\vhat_n$.
Suppose one has computed the solution
up to time $t_{n+3}$ inclusively. Using the solution computed in the 
last eight time steps, do the following.
First, find the average at $t=t_n$:
\bsube
\be
\widehat{\overline{v}}_n = 
\frac{11}{64} \vhat_n + \frac{15}{64}(\vhat_{n-1}+\vhat_{n+1}) - 
\frac{3}{32}(\vhat_{n-2}+\vhat_{n+2}) + \frac{1}{64}(\vhat_{n-3}+\vhat_{n+3}).  
\label{appnum_06a}
\ee
One can verify that $|\vhat_n - \widehat{\overline{v}}_n| = O(\dt^6)$. 
Next, repeat this for $t=t_{n-1}$:
\be
\widehat{\overline{v}}_{n-1} = 
\frac{11}{64} \vhat_{n-1} + \frac{15}{64}(\vhat_{n-2}+\vhat_{n}) - 
\frac{3}{32}(\vhat_{n-3}+\vhat_{n+1}) + \frac{1}{64}(\vhat_{n-4}+\vhat_{n+2}).  
\label{appnum_06b}
\ee
\label{appnum_06}
\esube
Finally, use $\widehat{\overline{v}}_n$ and $\widehat{\overline{v}}_{n-1}$
to restart the IF-LF method \eqref{appnum_05} by replacing, on the l.h.s.,
$\vhat_{n-1}$ with $\widehat{\overline{v}}_{n-1}$ and, on the r.h.s.,
$\vhat_{n}$ with $\widehat{\overline{v}}_{n}$. 
As demonstrated in \cite{2016JSci..72.14L},
this procedure suppresses the numerical instability while introducing only 
negligible dissipation to the solution. 
A logical flow-chart for implementing this stabilizing averaging in Matlab
is found at:\\ 
\verb+http://www.cems.uvm.edu/~tlakoba/recent_publications/stabilization_step_Eq28+\\
\verb+_logical_flowchart.txt+.

The IF-LF method with a stabilization step based on \eqref{appnum_06} needs
to use a time step $\dt$ satisfying
\be
\dt < \dx^2/\pi \equiv \pi/k_{\max}^2
\label{appnum_07}
\ee
in order to guarantee numerical stability of high-$k$ Fourier harmonics
\cite{2016JSci..72.14L}. In simulations reported in this study, 
we used the spatial
domain of length $L=40\pi$ and $N=2^{13}$ equally spaced grid points; this
corresponds to $\dx\approx 0.0153$ and $k_{\max}\approx 205$. According
to \eqref{appnum_07}, the threshold for high-$k$ numerical stability is
$\dt_{\rm thresh}=7.5\times 10^{-5}$, and we used $\dt=5\times 10^{-5}$ 
in all simulations
with the above $L$ and $N$. 
Whenever we needed to use different values of $L$ and $N$, as explained
in the text, we adjusted $\dt$ according to \eqref{appnum_07}.


\section*{Appendix B: Appropriateness of using periodic boundary conditions
 to simulate emergence of an intense pulse}

The reader may wonder whether the formation of an intense pulse,
reported in section 6, is not a numerical artefact caused by 
the imposed periodic boundary conditions.
Indeed, can the fact that the field passes through, and hence ``sees", 
the same computational domain multiple
times lead to an unphysical amplification of some of the field's
components? Here we present evidence that this is not the case.

\begin{figure} 
\begin{minipage}{7.5cm}
\hspace*{-0.1cm} 
\includegraphics[height=5.6cm,width=7.5cm,angle=0]{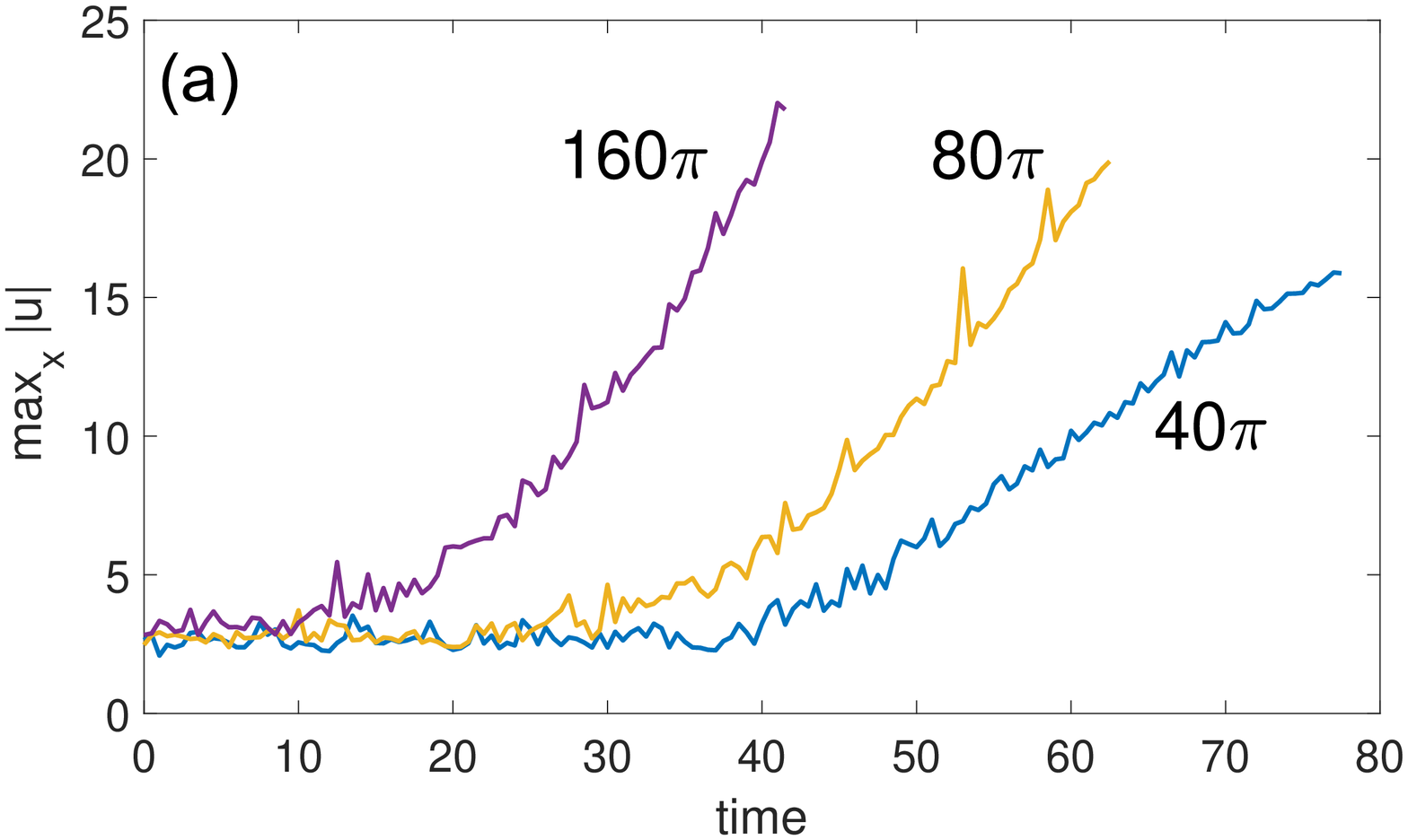}
\end{minipage}
\hspace{0.5cm}
\begin{minipage}{7.5cm}
\hspace*{-0.1cm} 
\includegraphics[height=5.6cm,width=7.5cm,angle=0]{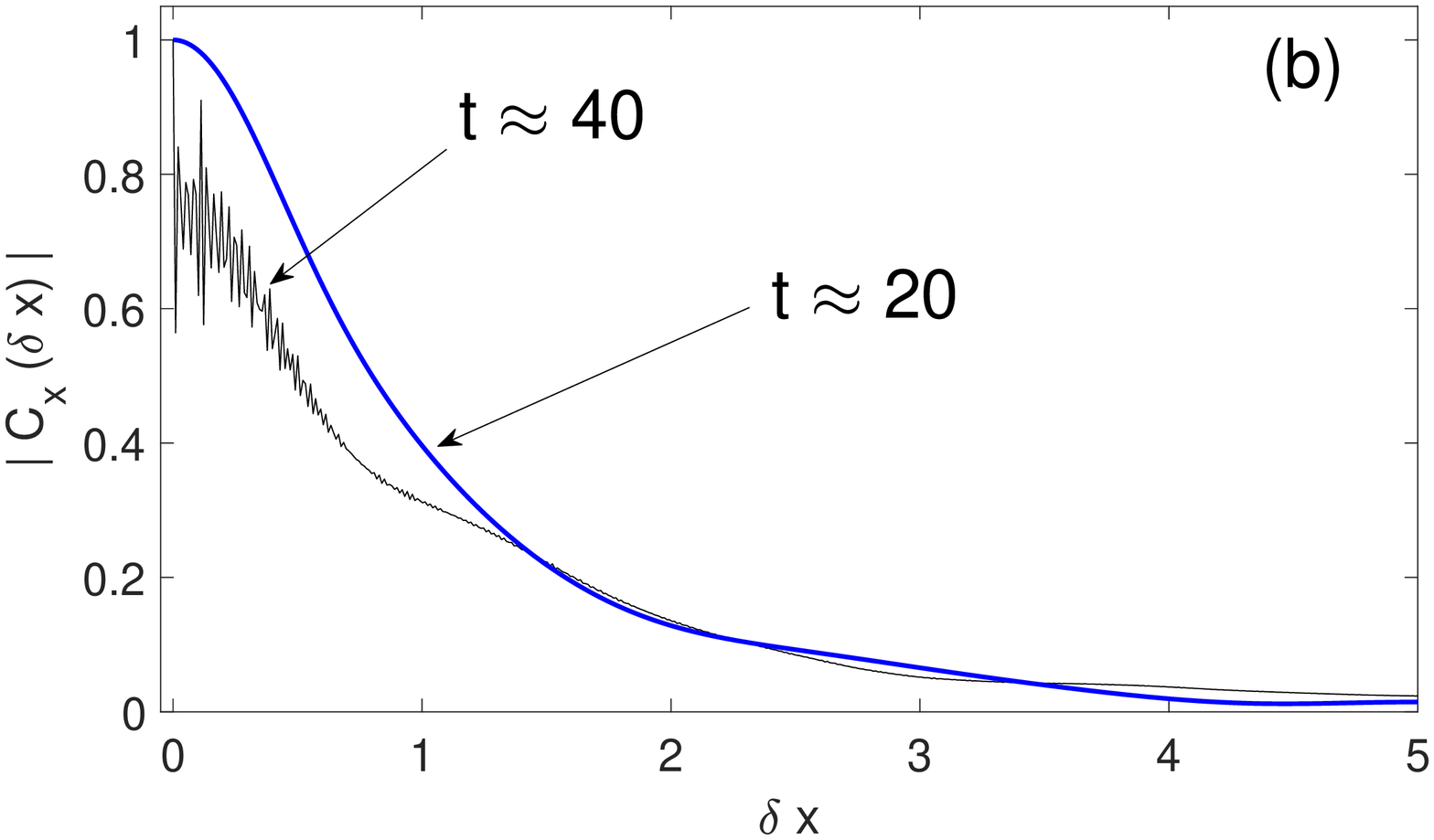}
\end{minipage}
\caption{(a) Evolution of maximum amplitude of the field for three
different lengths of the computational domain, listed in the figure.
 \ (b) Magnitude of the spatial correlation function of the field
for $L=160\pi$; see text for more details. 
}
\label{fig_Revision}
\end{figure}

First, despite the periodic boundary conditions, the field does not
have any considerable periodic component. We verified this by computing
the spatial correlation function 
\be
C_x(\delta x)=\langle u^*(x,t)u(x+\delta x,t)\rangle \,/\, 
              \langle |u(x,t)|^2 \rangle\,,
\label{appnum_B01}
\ee
where the angular brackets denote 
averaging that is performed both over $x$ at a given time as well
as over several time instances; see details in \cite{2016JSci..72.14L}.
The so computed spatial correlation of the field was found to decay
over distances $\delta x\sim 5$ (see Fig.~\ref{fig_Revision}(b) and
its description below), which is much smaller than length $L$ of
the computational domain. Thus, despite ``visiting" the same locations
of the computational domain multiple times, the field (and the emerging
pulse in particular) ``sees" a completely different environment every time.

Second, if, hypothetically, the periodicity of the computational domain
had played any constructive role in the pulse formation, then this formation
would be impeded by increasing the domain's length (since the longer
the computational domain, the more it is similar to a true infinite domain). 
Therefore, to test this hypothesis, 
we repeated the simulation reported in 
Fig.~\ref{fig_t2b}, except that we used two longer computational domains:
$L=80\pi$ and $L=160\pi$. The number of grid points was increased 
proportionally, so that the mesh size $\Delta x$ and hence the 
time step $\Delta t$ remained the same (see end of Appendix A).

In Fig.~\ref{fig_Revision}(a) we plot the maximum (over $x$) amplitude 
of the field for three different values of the domain length $L$. 
One can see that pulse formation is facilitated, rather than impeded,
by an increased domain. This is manifested in 
both that the pulse forms sooner
and that it reaches a greater amplitude in a longer domain. While a
rigorous explanation of these facts is unknown to us, the following are
plausible reasons behind them. First, 
an intense pulse begins to form out of
some part of the field that already has an above-average amplitude.
It seems likely that there are more locations with a stronger field in
a longer domain, and hence the pulse can begin to form sooner. 
Second, the pulse can reach a greater amplitude in a longer domain 
because it has the opportunity to ``harvest" energy from a larger
``reservoir", which is the area where the field has about-average
magnitude.\footnote{
  Note that due to periodic boundary conditions, the domain can be
	viewed as infinite. However, the amount of electromagnetic energy,
	available for ``harvesting" by the intense pulse,
	is still proportional to the computational length $L$. 
}

Finally, we return to our argument made after \eqref{appnum_B01} 
that the correlation length of the field is much smaller than the 
domain length. To refute the possibility that the periodicity of the
field (with period $L$) can affect pulse formation, 
it seems to make the most sense to compare $L$ with the field's
correlation length at a {\em time when the formation commences}. 
For the results shown in Fig.~\ref{fig_Revision}(a), these times
are approximately 40, 30, and 20 for $L=40\pi$, $80\pi$,
and $160\pi$, respectively. In Fig.~\ref{fig_Revision}(b) we show
the correlation function \eqref{appnum_B01} for $L=160\pi$ near
$t=20$; for completeness we also show the same function near $t=40$,
where the computation is terminated. The correlation functions
for $L=40\pi$ and $80\pi$ look qualitatively similar and therefore are
not shown. It is clear from Fig.~\ref{fig_Revision}(b) that the
field become decorrelated over distances that are more than an 
order of magnitude smaller than the length of the computational domain.


\section*{Appendix C: \ List of notations and acronyms used throughout the text}

\hspace*{0.55cm}
$\alpha=(+)$ or $(-)$: \ labels positrons or electrons, respectively, in secondary plasma.

$B_d$: \ dipolar component of magnetic field.

$B_s$: \ magnetic field on surface of pulsar.

$b=B_s/B_d$.

$c$: \ speed of light in vacuum.

$\Delta\gamma=|\gamma_{s,\,(+)}-\gamma_{s,\,(-)}|$: \ difference between mean Lorentz
   factors of positrons and electrons in secondary plasma.

$\Delta R$: \ distance along magnetic field lines where Langmuir turbulence is thought
   to be strong enough to lead to formation of coherent structures in charge density. 

$\Delta V$: \ potential drop across vacuum gap. 

$E_{\|}$: \ envelope of Langmuir wave. 

$E_0$: \ typical magnitude of $E_{\|}$ at a distance $R_{\rm onset}$. 

$e$: \ charge of electron ($e>0$).

$f_{\alpha}$: \ momentum distribution function of type $\alpha$ particles in secondary plasma.

$\gamma_b\sim 2\times 10^6$: average Lorentz factor of primary plasma beam particles.

$\gamma_p\sim 10^2$: average Lorentz factor of electrons and positrons in secondary plasma; \
    $\gamma_2=\gamma_p/10^2$.

$\gamma_{\alpha}, \,\gamma_T$: \ mean and standard deviation of Lorentz factor $\gamma$
       in the secondary plasma of type $\alpha$ particles.
			
$h\gtrsim 10^3$ cm: \ height of the vacuum gap above the polar cap.

$G$: \ group velocity dispersion coefficient in NLSE \eqref{NLSwdamp}.

$G_d$: \ related to $G$ by \eqref{pareq1}.

$\kappa=n_p/n_b \sim 10^4$; \ $\kappa_4=\kappa/10^4$.

$\lambda_l$: \ Langmuir wavelength in OFR. 

$\lambda_p$: \ Langmuir wavelength in MFR at some reference location (around $R\sim 50 R_s$). 
	
$L$: \ nondimensional length of the computational domain used in section 6. 

$l=c/\omega_p$: \ characteristic scale of Lamgmuir waves in MFR, used for normalization
   in section 5. 

$l_{\rm corr}$: \ nondimensional correlation length of random field in 
   initial condition \eqref{t5_01a} for Langmuir wave. 

$m_e$: \ mass of electron. 

MFR: \ moving frame of reference.

$n_GJ$: \ Goldreich--Julian density.

$n_b,\,n_p$: \ density of electron-positron pairs in primary and secondary plasma, respectively.

NLSE: \ nonlinear Schr\"odinger equation.

$\Omega = 2\pi/P$: \ angular frequency of pulsar (nondimensional). 

$\omega_l\sim 4\times 10^{11}$ s$^{-1}$: \ Langmuir frequency (in OFR).

$\omega_p \approx \omega_l/\gamma_p \sim 4\times 10^9$ s$^{-1}$: \ 
Langmuir frequency (in MFR)
  at some reference location (around $R\sim 50 R_s$). 

OFR: \ observer frame of reference. 

$P\sim 1$: \ period of pulsar rotation (in seconds).

$\dot{P}$: \ rate of pulsar slow-down (nondimensional); \ $\dot{P}_{-15}=\dot{P}/10^{-15}$.

$p_{\alpha}, \,p_T$: \ mean and standard deviation of the momentum distribution function
      $f_{\alpha}$ in the secondary plasma.

$q$: \ dimensional nonlinearity coefficient in NLSE \eqref{NLSwdamp}.

$q_d$: \ related to $q$ by \eqref{pareq2}.

$Q$: \ nondimensional nonlinearity coefficient in NLSE \eqref{tn_04}.

$\rho\sim 10^6$ cm: \ curvature radius of magnetic field lines in the vacuum gap; \ 
   $\rho_6=\rho/(10^6\,{\rm cm})$. 

$R_s$: \ radius of pulsar (assumed to be 10 km in this paper).

$R$: \ distance from pulsar; \ $R_{50}=R/(50R_s)\sim 1$. 

$R_{em}$: \ distance from the pulsar where coherent radio emission takes place. 

$R_{\rm onset}\sim 200$ km: \  \ distance from the pulsar where Langmuir turbulence is thought
   to begin to develop. 

$r$: \ relative part of random field in initial condition \eqref{t5_01a} for
    Langmuir wave. 

$s$: \ coefficient of nonlinear Landau damping in NLSE \eqref{NLSwdamp}.

$(s/q)$: \ magnitude of nonlinear Landau damping relative to purely cubic nonlinearity 
    in NLSE.

$s_d$: \ related to $s$ by \eqref{pareq3}.

$\theta$: \ ratio of the characteristic scales of the envelope and carrier 
   of the Langmuir wave,
   defined after \eqref{tn_01}; \ $\theta\sim 10^3\ldots 10^4$.

$\tau,\,\tau_{\circ}$: \ dimensional time variables in MFR and OFR, respectively; their
   relation is given after \eqref{NLSwdamp}. 
	
$t$: \ nondimensional time in MFR, related to $\tau$ by 	\eqref{tn_03}. 

$t_{4\times}$: \ time that it takes amplitude of intense pulse to exceed four 
    times average amplitude of initial state; see section 6.2. 

$u$: \ nondimensional envelope of Langmuir wave; see \eqref{tn_02}.

$\chi,\,\chi_{\circ}$: \ dimensional spatial variables along magnetic field lines
    in MFR and OFR, respectively; their
   relation is given after \eqref{NLSwdamp}. 
	
$x$: \ nondimensional space variable in MFR, related to $\chi$ by 	\eqref{tn_01}.

\end{document}